\documentclass[aps,pra,superscriptaddress,twocolumn,reprint, a4paper,noeprint]{revtex4-1}
\usepackage[utf8]{inputenc}
\usepackage{amsmath}
\usepackage{amsfonts}
\usepackage{amssymb}
\usepackage{dsfont}
\usepackage[usenames,dvipsnames,svgnames,table]{xcolor}
\usepackage[USenglish]{babel}
\usepackage{hyperref}
\usepackage{bm}
\usepackage{braket}
\usepackage{graphicx}

\allowdisplaybreaks
\DeclareMathOperator{\Tr}{Tr}
\DeclareMathOperator{\Nabla}{\mbox{\boldmath$\nabla$}}
\newcommand{\mspin}{\widehat}
\newcommand{\vb}[1]{\bm{#1}}

\newcommand{\anticomm}[1]{\big\{#1\big\}}
\newcommand{\bracketr}[1]{\left( #1 \right)}
\newcommand{\bracket}[1]{\left[ #1 \right]}

\newcommand{\mat}[1]{\underline{#1}}

\newcommand{\rate}[2]{I_{#1}^{#2}}
\newcommand{\ratec}[2]{\vb{I}_{#1}^{#2}}

\newcommand{\ratem}[2]{\mat{I}_{#1}^{#2}}

\newcommand{\sig}{\widehat{\vb{\sigma}}}
\newcommand{\signull}{\widehat{\sigma}_0}
\newcommand{\one}{\mathds{1}}
\newcommand{\Snull}{\widehat{S}_0}
\newcommand{\Sx}{\widehat{S}_1}
\newcommand{\Sy}{\widehat{S}_2}
\newcommand{\Sz}{\widehat{S}_3}
\newcommand{\Qx}{\widehat{Q}_1}
\newcommand{\Qy}{\widehat{Q}_2}
\newcommand{\Qz}{\widehat{Q}_3}
\newcommand{\Qxy}{\widehat{Q}_4}
\newcommand{\Qzr}{\widehat{Q}_5}
\newcommand{\Sp}{\vb{\widehat{S}}}
\newcommand{\Q}{{\widehat{Q}}}

\newcommand{\dep}{(\vb{r}, \vb{k}, t)}
\newcommand{\depend}{\gs, \electron, \hole, \triplet}

\newcommand{\spin}[2]{\widehat{#1}_{#2}}

\newcommand{\psing}{\widehat{P}_s}
\newcommand{\ptrip}{\widehat{P}_t}
\newcommand{\pdoub}{\widehat{P}_d}

\newcommand{\gs}{f_s}

\newcommand{\gsc}{f''_s}

\newcommand{\feup}{f_{e \uparrow}}
\newcommand{\fedown}{f_{e \downarrow}}
\newcommand{\feupa}{f'_{e \uparrow}}
\newcommand{\fedowna}{f'_{e \downarrow}}
\newcommand{\feupc}{f''_{e \uparrow}}
\newcommand{\fedownc}{f''_{e \downarrow}}

\newcommand{\electron}{\spin{{f}}{e}}
\newcommand{\me}{n_{e}}

\newcommand{\ce}{\vb{c}_e}

\newcommand{\electrona}{\widehat{f}'_{e}}
\newcommand{\mea}{n'_{e}}

\newcommand{\ceza}{c_{e}^{z\prime}}
\newcommand{\cea}{\vb{c}'_e}

\newcommand{\electronc}{\widehat{f}''_{e}}
\newcommand{\mec}{n''_{e}}

\newcommand{\cec}{\vb{c}''_e}

\newcommand{\elec}{\spin{{f}}{e0}}
\newcommand{\elecg}{\spin{{g}}{e}}

\newcommand{\fhupa}{f'_{h \uparrow}}
\newcommand{\fhdowna}{f'_{h \downarrow}}

\newcommand{\fhupd}{f'''_{h \uparrow}}
\newcommand{\fhdownd}{f'''_{h \downarrow}}

\newcommand{\hole}{\spin{{f}}{h}}
\newcommand{\mh}{n_{h}}

\newcommand{\ch}{\vb{c}_h}

\newcommand{\holea}{\widehat{f}'_{h}}
\newcommand{\mha}{n'_{h}}

\newcommand{\cha}{\vb{c}'_h}

\newcommand{\holed}{\widehat{f}'''_{h}}
\newcommand{\mhd}{n'''_{h}}

\newcommand{\chzd}{c_{h}^{z\prime\prime\prime}}
\newcommand{\chd}{\vb{c}'''_h}

\newcommand{\holee}{\spin{{f}}{h0}}
\newcommand{\holeg}{\spin{{g}}{h}}

\newcommand{\mhe}{n_{h0}}
\newcommand{\che}{\vb{c}_{h0}}
\newcommand{\mhg}{\tilde{n}_h}
\newcommand{\chg}{\tilde{\vb{c}}_h}
\newcommand{\chgi}{\tilde{c}_{hi}}

\newcommand{\triplet}{\spin{{f}}{t}}
\newcommand{\mto}{{n_{t}}}

\newcommand{\ctz}{{c_{t}^z}}
\newcommand{\ct}{\vb{c}_t}
\newcommand{\qtx}{{q_{t1}}}
\newcommand{\qty}{{q_{t2}}}
\newcommand{\qtz}{{q_{t3}}}
\newcommand{\qtxy}{{q_{t4}}}
\newcommand{\qtzr}{{q_{t5}}}

\newcommand{\tripletb}{\spin{{f}}{t}}
\newcommand{\mtob}{n_{t}}

\newcommand{\qt}{\vec{q}_t}

\newcommand{\qeh}{\vec{q}_{eh}}

\newcommand{\qtb}{\vec{q}_t}
\newcommand{\matq}{\mat{q}{}_t}
\newcommand{\ctb}{\vb{c}_{t}}
\newcommand{\matqb}{\mat{q}{}_t}

\newcommand{\tripleta}{\widehat{f}'_{t}}
\newcommand{\mtoa}{n'_{t}}
\newcommand{\cta}{\vb{c}'_{t}}
\newcommand{\matqa}{\mat{q}'{\hspace{-0.3em}}_t}

\newcommand{\ctza}{c_t^{z\prime}}

\newcommand{\qtzra}{q'_{t5}}
\newcommand{\qta}{\vec{q}^{\:\prime}_{t}}

\newcommand{\tripletc}{\widehat{f}''_{t}}
\newcommand{\mtoc}{n''_{t}}
\newcommand{\mtott}{{n}_{tt}}
\newcommand{\ctc}{\vb{c}''_{t}}
\newcommand{\matqc}{\mat{q}''{\hspace{-0.5em}}_t\hspace{0.15em}}

\newcommand{\qtc}{\vec{q}^{\:\prime\prime}_{t}}

\newcommand{\tripletd}{\widehat{f}'''_{t}}
\newcommand{\mtod}{n'''_t}

\newcommand{\qtzrd}{q'''_{t5}}
\newcommand{\ctd}{\vb{c}'''_{t}}
\newcommand{\matqd}{\mat{q}'''_t}

\newcommand{\ctzd}{c_t^{z\prime\prime\prime}}
\newcommand{\qtd}{\vec{q}^{\:\prime\prime\prime}_{t}}

\newcommand{\ftp}{f_{t,1}}
\newcommand{\ftm}{f_{t, \bar{1}}}
\newcommand{\ftz}{f_{t,0}}

\newcommand{\tp}{t_{1}}
\newcommand{\tm}{t_{\bar{1}}}
\newcommand{\tz}{t_{0}}

\newcommand{\ftpd}{f'''_{t, 1}}
\newcommand{\ftmd}{f'''_{t, \bar{1}}}
\newcommand{\ftzd}{f'''_{t, 0}}

\newcommand{\ftpa}{f'_{t, 1}}
\newcommand{\ftma}{f'_{t, \bar{1}}}
\newcommand{\ftza}{f'_{t, 0}}

\newcommand{\ftpb}{f_{t, 1}}
\newcommand{\ftmb}{f_{t, \bar{1}}}
\newcommand{\ftzb}{f_{t, 0}}

\newcommand{\ftpc}{f''_{t, 1}}
\newcommand{\ftmc}{f''_{t, \bar{1}}}
\newcommand{\ftzc}{f''_{t, 0}}

\newcommand{\ctzn}{c_{t0}^{z}}

\newcommand{\qtt}{\vec{q}_{tt}}
\newcommand{\ctt}{\vb{c}_{tt}}
\newcommand{\matqtt}{\mat{q}{}_{tt}}

\newcommand{\mtoeh}{n_{eh}}

\newcommand{\intermed}{\widehat{{f}}_{eh}}
\newcommand{\ceh}{\vb{c}_{eh}}
\newcommand{\matqeh}{\mat{q}{}_{eh}}

\newcommand{\intermedinv}{\widehat{{f}}_{1 - eh}}

\newcommand{\deviat}[1]{\tilde{#1}}

\newcommand{\scatpart}[2]{\widehat{I}_{#2}\big[ #1 \big]}
\newcommand{\scatsing}[1]{I_{s}\big[ #1 \big]}

\newcommand{\sigx}{\widehat{\sigma}_x}
\newcommand{\sigy}{\widehat{\sigma}_y}
\newcommand{\sigz}{\widehat{\sigma}_z}
\newcommand{\sigi}{\widehat{\sigma}_i}

\newcommand{\meo}{n_{e0}}
\newcommand{\ceo}{c_{e0}}
\newcommand{\meg}{\tilde{n}_{e}}

\newcommand{\ceg}{\tilde{\vb{c}}_e}
\newcommand{\cegi}{\tilde{c}_{ei}}

\newcommand{\singf}{f_{s}}
\newcommand{\singfo}{f_{s0}}
\newcommand{\singg}{g_{s}}

\newcommand{\tripletfo}{\widehat{f}_{t0}}
\newcommand{\tripletg}{\widehat{g}_{t}}
\newcommand{\qtv}{\vec{q}_t}
\newcommand{\gmat}{\mat{\widehat{Q}}}
\newcommand{\gmatij}{\widehat{Q}}
\newcommand{\mtoo}{{n_{t0}}}
\newcommand{\cto}{{c}_{t0}}
\newcommand{\mtog}{\tilde{n}_t}
\newcommand{\ctig}{\tilde{c}_{ti}}
\newcommand{\ctkg}{\tilde{c}_{tk}}
\newcommand{\ctg}{\tilde{\vb{c}}_t}
\newcommand{\ctxg}{\tilde{c}_t^x}
\newcommand{\ctyg}{\tilde{c}_t^y}
\newcommand{\ctzg}{\tilde{c}_t^z}
\newcommand{\qtvg}{\tilde{\vec{q}}_t}
\newcommand{\qtxg}{\tilde{q}_{t1}}
\newcommand{\qtyg}{\tilde{q}_{t2}}
\newcommand{\qtzg}{\tilde{q}_{t3}}
\newcommand{\qtxyg}{\tilde{q}_{t4}}
\newcommand{\qtzrg}{\tilde{q}_{t5}}

\newcommand{\atg}{\deviat{\vb{a}}}
\newcommand{\btg}{\deviat{\vb{b}}}

\newcommand{\atgi}{\deviat{{a}}}
\newcommand{\btgi}{\deviat{{b}}}

\begin{document}

\title{Spin-conserving Boltzmann theory for carriers and excitons in organic semiconductors}

\author{Charlotte B\"acker}
\affiliation{Institute of Theoretical Physics, Technische Universit\"at Dresden, 01062 Dresden, Germany}

\author{Linus Thummel}  
\affiliation{Imperial College London, South Kensington Campus, London SW7 2AZ, United Kingdom} 
\affiliation{Institute of Theoretical Physics, Technische Universit\"at Dresden, 01062 Dresden, Germany}

\author{Carsten Timm}
\email{carsten.timm@tu-dresden.de}
\affiliation{Institute of Theoretical Physics, Technische Universit\"at Dresden, 01062 Dresden, Germany}
\affiliation{Würzburg-Dresden Cluster of Excellence ct.qmat, Technische Universit\"at Dresden, 01062 Dresden, Germany}

\date{August 6, 2021}

\begin{abstract}
The rise of organic electronics calls for versatile modeling tools. In this context, we develop a semiclassical Boltzmann theory that describes transport and excitonic processes in crystalline organic semiconductors on equal footing. The generation of singlet and triplet excitons out of the ground state, their formation from free electrons and holes, the reverse processes, as well as the fusion and fission of excitons are included. The corresponding scattering integrals respect spin conservation, which requires matrix-valued distribution functions. They also include fermionic and bosonic many-particle effects such as Pauli blocking. We employ a multipole expansion of the distribution functions, where quadrupolar terms turn out to be essential for the triplet excitons. This work provides a basis for the modeling of organic solar cells, in which excitonic processes are crucial for the performance. Moreover, the theory is of general interest for transport and transitions of multiple (quasi-) particle species carrying spin in nonequilibrium systems.
\end{abstract}

\maketitle

\section{Introduction}
\label{sec.introduction}

Organic semiconductors are promising materials for nanoelectronics and photonics. Their phy\-si\-co\-che\-mi\-cal properties are easily adjustable by chemical engineering, the production process is inexpensive, and they are advantageous in situations where low weight or mechanical flexibility are desired \cite{For04,Muc06,Kla06,BrA12,Sir14,RSP17}. Moreover, they are reaching conductivities comparable to those of metals~\cite{CFP77,BCF02,PMB04,JBP04,KGM15,MaM17}.

However, the design of electronic properties requires a deeper theoretical understanding of the charge-car\-rier-trans\-port mechanisms. Several aspects make this a challenging goal for organic materials: On the one hand, the lattice is rather soft, i.e., the predominantly intermolecular phononic modes have low energies. They are also relatively strongly coupled to the charge carriers. This interaction is of short range with the largest contribution being the coupling of the electronic occupation of a certain molecular orbital to vibrations of the same molecule. This diagonal or \emph{Holstein} coupling \cite{Hol59} results in the formation of so-called small polarons \cite{Mah90,note.smallpolaron}. There is also a sizable coupling of intermolecular tunneling amplitudes to vibrations, called nondiagonal or \emph{Peierls} coupling \cite{Gla63,GoC66,SSH79,MSB17,FGB20}. For rubrene, Ordej\'on \textit{et al.}\ \cite{OBP17} have found that the Peierls coupling is indeed important. These couplings lead to strong polaronic effects \cite{OBH09,OrR11,CiF12,FCM17,FGB20,HPO21}, which limit the carrier mobility.

On the other hand, creation, decay, and recombination of excitons as well as the fission of spin-singlet excitons into triplet excitons and the reverse fusion process are crucial for organic photovoltaic and light-emitting devices \cite{RWL07,WRL10,SmM10,Sil10,HRG12,MLH13}: Lifetimes of triplet excitons can be six orders of magnitude longer than the ones of singlet excitons because their transition into the ground state, i.e., the recombination process, requires a spin flip \cite{Mik12}. Due to the spin exchange energy, the singlet-exciton energy can be higher than twice the triplet-exciton energy \cite{MoF15}. This enables singlet fission into two triplet excitons, which is spin allowed. This fission process can be the key mechanism to induce transport of excitation energy by triplet excitons over long distances, which is of major importance for photovoltaic applications~\cite{SmM10,Sil10}. Note that the photogeneration in such devices is dominated by internal interfaces while the exciton diffusion is a bulk effect. We return to this point in the conclusions.

We aim at a unified description of transport and the aforementioned local excitonic processes. A relevant property of organic materials is the typically weak spin-orbit coupling, which ensures that spin is conserved to a good approximation. The description of transport thus has to respect spin rotation symmetry, in particular if singlet and triplet excitons are involved. We note that any violation of spin conservation, i.e., any relaxation of spin, can easily be incorporated by an additional scattering integral.

The excitons in typical devices are far from equilibrium. The quantum master equation constitutes a powerful tool for the description of nonequilibrium processes in quantum systems \cite{BrP02}. It is the equation of motion for the reduced density matrix or statistical operator $\widehat\rho$ of an open system. This equation is linear in the density matrix since it is derived from the Schr\"odinger equation, which is linear in the state vector. Linearity is necessary to preserve the unit trace of the density matrix. However, one can relax this trace condition and interpret $\Tr\widehat\rho$ as a concentration of quasiparticles. The matrix structure of $\widehat\rho$ then encodes the distribution of spin orientations, where $\widehat\rho$ is a $(2s+1)\times (2s+1)$ matrix for spin-$s$ quasiparticles. Furthermore, the equation of motion can then be nonlinear, for example describing generation and decay processes. A generalized quantum master equation in this sense was applied to local excitonic processes, such as fission and fusion, in organic crystals as early as 1969 by Johnson and Merrifield \cite{JoM70}. Essentially the same equation was proposed, under the name of \emph{stochastic Liouville equation}, in the context of quantum optics \cite{ScL67} and for the description of chemical reactions involving spin \cite{EFL73,Hab76,StU89}. More recently, Schellekens \textit{et al.}\ \cite{SWK11} have applied it to excitonic processes in order to describe organic magnetoresistance.

In case of the stochastic Liouville equation, the nonlinearity is given by a $\widehat\rho$-independent source term \cite{JoM70,ScL67,SWK11}. The equation takes the form
\begin{equation}
\frac{\partial\widehat\rho}{\partial t} = - \frac{i}{\hbar}\, \big[\widehat H,\widehat\rho\big]
  - \frac{W}{2}\, \big\{ \widehat P, \widehat\rho \big\} + \widehat\Gamma .
\end{equation}
The first term on the right-hand side describes the unitary time evolution determined by the Hamiltonian $\widehat H$. The second term implements decay processes and is naturally linear in the density. Here $W$ is a decay rate and $\widehat P$ is a projection operator onto the spin state from which a decay is allowed by spin conservation. The specific anticommutator structure was shown by Haberkorn \cite{Hab76} to ensure the positivity of $\widehat\rho$---since the eigenvalues of $\widehat\rho$ are interpreted as concentrations they must not become negative \cite{note.Haberkorn}. The third term describes generation processes and is of order zero in $\widehat\rho$ since the generation rate is assumed to be independent of the density.

The interpretation of $\widehat\rho$ as a density also permits terms of higher order. Consider, for example, a pair-an\-ni\-hi\-la\-tion process of the type $A+A\to 0$. For spinless particles, $\widehat\rho$ just becomes the number density $n$ and the process is naturally described by the equation
\begin{equation}
\frac{\partial n}{\partial t} = - c\, n^2 .
\end{equation}
This is of course well known in chemical kinetics \cite{McQ67,nonlinear}. Nonlinear terms also result from quantum statistics, which show that the rate of a scattering process also depends on the occupation of the final state, e.g., due to the Pauli principle.

The question arises what the corresponding equation looks like for particles with spin. The right-hand side then contains a product of the matrix $\widehat\rho$ with itself and the question is how this product must be constructed so that spin is conserved. More complicated processes involving multiple species, e.g., $A+B\to C$, can be described using multiple matrix-valued densities and we have to construct spin-conserving terms out of matrices of generally different sizes $(2s+1)\times(2s+1)$. This issue is central for the description of excitonic processes. For spin-$1/2$ quasiparticles, i.e., electrons and holes, the matrix-valued densities are $2\times 2$ matrices acting on their spin Hilbert space. For triplet excitons with spin $1$, they are $3\times 3$ matrices, whereas for singlet excitons they reduce to scalars.

We combine the spin-rotation-invariant description of local excitonic processes with a spin-rotation-invariant description of carrier and exciton transport. We are interested in bandlike transport in clean organic materials at not very low temperatures. Characteristic quantum effects such as weak localization and universal conductance fluctuations are thus expected to be irrelevant and a semiclassical description is justified. Our goals thus require the derivation of $\mathrm{SU}(2)$ spin-rotation-invariant Boltzmann-type kinetic equations for the various quasiparticle species and of scattering integrals describing transitions between them.

As noted above, polaronic effects are typically strong in organic materials \cite{OBH09,OrR11,CiF12,FCM17,FGB20,HPO21}. The coupling to the phonons causes the charge carriers and also excitons to dress as small polarons \cite{note.smallpolaron}. Unlike for inorganic materials, there is no well-developed semiclassical transport theory for small polarons in organic semiconductors. This is a promising topic for future research, where one obstacle will likely be the inclusion of sizable Peierls coupling. Numerical results by Hutsch \textit{et al.}\ \cite{HPO21} indicate that the phononic modes can be usefully separated into fast modes, which lead to polaron formation similar to inorganic materials, and slow modes, which can be treated as quasistatic disorder. This suggests that a semiclassical description should be possible, where the effect of the slow modes is incorporated by scattering integrals. Since polaronic effects are not at the focus of the present paper we assume that the polarons are adiabatically connected to the undressed quasiparticles, i.e., that polaronic effects only renormalize the parameters appearing in our description but do not change its structure.

Spin-rotation invariance requires the Boltzmann theory to contain the matrix-valued densities \cite{SmJ89,SXW97,Sim01,QiZ03,ZLZ04,PiT07,GSR10,PoN11,Haj14}. We emphasize that it is not sufficient to use kinetic equations for the densities in each spin channel, say spin-up and spin-down electrons, since this is equivalent to assuming the matrix-valued densities to be diagonal in the standard spin basis. This precludes the description of spin polarization in the \textit{x} and \textit{y} directions. It is also clear that the scattering integrals are significantly more complicated than for spinless particles since they depend on matrix-valued functions of various dimensions and are themselves matrices.

Our program is of general interest beyond the field of organic electronics as the principles developed here are useful for any system showing scattering of and reactions between spin-carrying particles, for example in spintronics and spin chemistry.

This paper is organized as follows. In Sec.\ \ref{sec.kineticeq}, we set up the spin-conversing kinetic equations for charge carriers and excitons in uniform electric and magnetic fields. The treatment of scattering and transitions between quasiparticle species (excitonic processes) is then performed in Sec.\ \ref{sec.scattering}. The theoretical framework is illustrated for a simple model in Sec.\ \ref{sec.example}. Conclusions are drawn in Sec.~\ref{sec.concl}.

\section{Spin-conserving kinetic equations}
\label{sec.kineticeq}

In this section, we present the Boltzmann transport equations for electrons, holes, and singlet and triplet excitons including static, uniform electric and magnetic fields. We concentrate on the drift terms conventionally appearing on the left-hand side of the Boltzmann equation and leave the collision terms for Sec.~\ref{sec.scattering}.

Following Ref.\ \cite{QiZ03}, we write the general Boltzmann equation for the density $\widehat f\dep$ as
\begin{align}
&\frac{\partial}{\partial t}\, \widehat f\dep
  + \vb{v}(\vb{k}) \cdot \Nabla_{\vb{r}} \widehat f\dep \nonumber \\
&\quad{}+ \frac{q}{\hbar} \left( \vb{E} + \vb{v}(\vb{k})\times\vb{B} \right)
    \cdot \Nabla_{\vb{k}} \widehat f\dep \nonumber \\
&\quad{}+ \frac{i}{\hbar} \big[ \widehat H\dep, \widehat f\dep \big]
  = \widehat{I}[\widehat f] ,
\label{2.Boltzmann.3}
\end{align}
where $q \in \{-e,0,e\}$ is the charge of the quasiparticles under consideration and the hat $\widehat\bullet$ denotes matrices acting on spin space. The density $\widehat f\dep$ is a $(2s+1)\times (2s+1)$ matrix, where $s$ is the quasiparticle spin. Note that for $s>0$, the matrix structure leads to an additional commutator term, which describes the dynamics in spin space \cite{QiZ03}. For example, for spin-$1/2$ quasiparticles in a magnetic field $\vb{B}$, the Hamiltonian $\widehat H$ contains a Zeeman term proportional to $\vb{B}\cdot\widehat{\mbox{\boldmath$\sigma$}}$, where $\widehat{\mbox{\boldmath$\sigma$}}$ is the vector of Pauli matrices $\widehat\sigma_1$, $\widehat\sigma_2$, $\widehat\sigma_3$. We suppress the arguments $\vb{r}$, $\vb{k}$, and $t$ from now on. $\widehat{I}[\widehat f]$ contains all scattering integrals affecting the distribution.

Since we will later consider nonlinear collision integrals, the normalization of $\widehat f$ is important. We take the eigenvalues of $\widehat f$ to be the occupation numbers per quantum state. For example, the density
\begin{equation}
\widehat f_e = \begin{pmatrix}
    1 & 0 \\
    0 & 1
  \end{pmatrix}
\end{equation}
for all $\vb{r}$ and $\vb{k}$ describes a filled electron band. This convention will be useful for the description of Fermi and Bose statistics in the collision integrals. As noted above, $\Tr\widehat f$ is not restricted to unity and $\widehat f$ is not a density matrix in the sense of a statistical operator~\cite{SWK11}.

The velocity in Eq.\ (\ref{2.Boltzmann.3}) is given by the derivative of the Hamiltonian acting on spin space with respect to momentum,
\begin{equation}
\vb{v} = \frac{1}{\hbar}\, \frac{\partial \widehat H}{\partial\vb{k}} ,
\label{eq:evelocity_matrix}
\end{equation}
and thus, in principle, obtains a matrix structure. This structure is made nontrivial by those terms in the Hamiltonian that contain both the momentum and the spin, i.e., by spin-orbit coupling. Since we are primarily interested in organic materials, which show weak spin-orbit coupling, we neglect them. The velocity $\vb{v}$ then becomes proportional to the identity matrix and can be treated as a scalar~\cite{note.soc}. The dispersion relation is the only relevant momentum-dependent term in the Hamiltonian so that we can express $\vb{v}$ as the group velocity 
\begin{equation}\label{eq:group vlocity}
	\vb{v} =  \frac{1}{\hbar} \vb{\nabla}_{\vb{k}} \epsilon(\vb{k})
\end{equation} 
in terms of the quasiparticle dispersion $\epsilon(\vb{k})$.

\subsection{Electrons}

The charge carriers in organic semiconductors are electrons and holes. For simplicity, we restrict ourselves to a model with one valence band and one conduction band; the generalization to multiple bands is straightforward. The spin $1/2$ of the electrons and holes is essential to understand the related excitonic processes.

The identity matrix $\signull$ and the three Pauli matrices $\sigx$, $\sigy$, and $\sigz$ form a basis of the space of Hermitian $2\times 2$ matrices. Hence, for spin-$1/2$ quasiparticles, the $2\times 2$ matrix $\widehat f$ can be expanded as
\begin{equation}
\widehat f = \frac{1}{2} \left( n\, \signull + \vb{c} \cdot \sig \right) .
\label{2a.f.multipole.3}
\end{equation}
Then $n = \Tr \widehat f$ is the concentration of quasiparticles, summed over spin orientations, and $\vb{c} = \Tr \widehat f\, \sig$ is twice the spin density (setting $\hbar=1$). Note that $\widehat f$ must be Hermitian to ensure real densities. We will call relations of the form of Eq.\ (\ref{2a.f.multipole.3}) \emph{multipole expansions}.

We start with the description of electrons in the conduction band with the Hamiltonian
\begin{equation}
	\widehat{H}_e  = \widehat{H}_{e0} + \frac{1}{2}\, g_\text{el} \mu_B \vb{B} \cdot \sig ,
	\label{eq:hamiltonian e}
\end{equation}
with the electronic Land\'e factor $g_\text{el}$ and the Bohr magneton $\mu_B$. Neglecting spin-orbit coupling, the bare electron Hamiltonian is $\widehat{H}_{e0} = \epsilon_e(\vb{k})\, \signull$, with the dispersion $\epsilon_e(\vb{k})$. The Zeeman term remains the only part with nontrivial matrix structure. 

We decompose the electronic distribution function into the equilibrium distribution $\elec$ and a deviation $\elecg$ according to 
\begin{equation}
	\electron = \elec + \elecg .
	\label{eq:linear response f_e}
\end{equation}
We assume the distribution function to be close to equilibrium because the equilibrium distribution is perturbed only little by (a) a weak electric field $\vb{E}$ and (b) decomposition of a low concentration of excitons. The weak electric field also means that terms involving the product of $\vb{E}$ and $\elecg$ are small of higher order and can be neglected. This is the assumption of linear response. It would be straightforward to drop this approximation.

The multipole expansion of the deviation is 
\begin{equation}
	\elecg = \frac{1}{2} \left( \meg \signull + \ceg \cdot \sig \right) ,
	\label{eq:g_e multipole}
\end{equation}
similar to Eq.\ (\ref{2a.f.multipole.3}). In equilibrium, the distribution is spatially uniform since the external fields and the scattering term are assumed not to depend on position $\vb{r}$, and can thus be written as
\begin{equation}
	\elec =  \frac{1}{2} \left( \meo\,\signull + \ceo \vb{B} \cdot \sig \right) .
	\label{eq:f_e0 multipole}
\end{equation}
This yields
\begin{equation}
  \electron = \frac{1}{2}\, \big[ (\meo+\meg)\,\signull + (\ceo \vb{B} + \ceg) \cdot \sig \big] .
  \label{eq:f_e multipole full}
\end{equation}
The equilibrium distribution can be derived starting from the specific case of $\vb{B} \parallel \vb{e}_3$, where $\vb{e}_3$ is the unit vector in the $z$ direction. Here the Hamiltonian takes a diagonal form and so does the equilibrium function. It is thus given by a linear combination of $\signull$ and $B\, \sigz$. The rotation to an arbitrary direction of the magnetic field is then described by the transformation $B\, \sigz \rightarrow \vb{B}\cdot\sig$. The induced magnetization is along $\vb{B}$ since we assume a magnetically isotropic medium \cite{note.nonlinear}. The equilibrium term $\meo$ corresponds to the Fermi-Dirac distribution of electrons with energy $\epsilon_e(\vb{k})$, chemical potential $\mu$, and temperature $T$,
\begin{equation} 
\meo(\vb{k}) = \frac{1}{e^{(\epsilon_e(\vb{k}) - \mu)/k_BT} + 1} .
	\label{eq:Fermi Dirac electrons}
\end{equation}
In the linear-response approximation, the Boltzmann equation for the electrons reads as
\begin{align}
	&\frac{\partial}{\partial t}\, \elecg + \frac{1}{\hbar}\, \vb{\nabla}_{\vb{k}}
	  \epsilon_e \cdot \vb{\nabla}_{\vb{r}} \elecg - \frac{e}{\hbar} \left( \vb{E}
	  + \frac{1}{\hbar}\, \vb{\nabla}_{\vb{k}} \epsilon_e \times \vb{B} \right)
	  \cdot \vb{\nabla}_{\vb{k}}\elec \nonumber\\
	& \quad{}- \frac{e}{\hbar^2} \left(\vb{\nabla}_{\vb{k}} \epsilon_e \times \vb{B}\right)
	  \cdot \vb{\nabla}_{\vb{k}} \elecg + \frac{i}{\hbar}\, \big[\mspin{H}_e, \elec \big]
	  + \frac{i}{\hbar}\, \big[\mspin{H}_e,\elecg \big] \nonumber\\
	&=2 \scatpart{\depend}{e} ,
	\label{eq:elec Boltzmann linear response}
\end{align}
where the factor of 2 is due to the factor of $1/2$ in Eq.\ (\ref{2a.f.multipole.3}) and drops out when the scattering term is also expanded. We now analyze the individual expressions in the multipole expansion. 
The derivative 
\begin{equation}
	\vb{\nabla}_{\vb{k}} \elec = \frac{\meo^2 - \meo}{2 k_B T}\,
	\vb{\nabla}_{\vb{k}} \epsilon_e \signull + \frac{1}{2}\, \vb{\nabla}_{\vb{k}} \ceo \vb{B} \cdot \sig
\end{equation}
can be used to simplify the expression
\begin{equation}
	\left(\vb{\nabla}_{\vb{k}} \epsilon_e \times \vb{B} \right) \cdot \vb{\nabla}_{\vb{k}} \elec
    = \frac{1}{2}\, (\vb{\nabla}_{\vb{k}} \epsilon_e \times \vb{B} )
    \cdot (\vb{\nabla}_{\vb{k}} \ceo) \vb{B} \cdot \sig .
\end{equation}
The commutator terms are
\begin{align}
	\frac{i}{\hbar} \big[\mspin{H}_e, \mspin{f}_{e0} \big] &= 0 , \\
	\frac{i}{\hbar} \big[\mspin{H}_e, \mspin{g}_e \big]
	   &= - \frac{g_\text{el} \mu_B}{2 \hbar}\, (\vb{B} \times \vb{c}_e)\cdot\sig ,
\end{align}
where the latter expression describes the precession of the spin around the direction of the magnetic field.

Inserting these expressions into the Boltzmann equation (\ref{eq:elec Boltzmann linear response}) gives 
\begin{align}
&\frac{\partial}{\partial t}\, \meg \signull + \frac{\partial}{\partial t}\, \ceg \cdot \sig \nonumber\\
& \quad{} + \frac{1}{\hbar}\, \vb{\nabla}_{\vb{k}} \epsilon_e \cdot
  \left[ \vb{\nabla}_{\vb{r}} \meg \signull + \vb{\nabla}_{\vb{r}} (\ceg \cdot \sig) \right] \nonumber\\
& \quad{} - \frac{e}{\hbar}\, \vb{E} \cdot
  \big[ \vb{\nabla}_{\vb{k}} \meo \signull + \vb{\nabla}_{\vb{k}} \ceo \vb{B} \cdot \sig \big] \nonumber\\
& \quad{} - \frac{e}{\hbar^2} \left( \vb{\nabla}_{\vb{k}} \epsilon_e \times \vb{B} \right) \cdot
  \vb{\nabla}_{\vb{k}} \ceo \vb{B} \cdot \sig \nonumber\\
& \quad{} - \frac{e}{\hbar^2} \left( \vb{\nabla}_{\vb{k}} \epsilon_e \times \vb{B} \right) 
  \cdot \left[ \vb{\nabla}_{\vb{k}} \meg \signull + \vb{\nabla}_{\vb{k}} (\ceg \cdot \sig) \right]
  \nonumber \\
& \quad{} - \frac{g_\text{el} \mu_B}{\hbar} \left(\vb{B} \times \vb{c}_e \right) \cdot \sig
  \nonumber \\
&= \scatpart{\depend}{e} . 
\label{eq:elec Boltzmann multipole}
\end{align}
It is useful to also expand the scattering term into multipole components, in analogy to Eq.\ (\ref{2a.f.multipole.3}),
\begin{align}
\widehat{I}_e &= \frac{1}{2} \left( I^{n_e}\, \signull + \vb{I}^{\vb{c}_e} \cdot \sig \right) \nonumber \\
&= \frac{1}{2} \left( I^{n_e}\, \signull
  + I^{c_e^x}\, \sigx + I^{c_e^y}\, \sigy + I^{c_e^z}\, \sigz \right) .
\label{eq:multipole scattering term}
\end{align}
From this, we can obtain equations for the coefficients of the basis matrices. The notation can be shortened using the differential operator
\begin{equation}
	\mathcal{D}_e = \frac{\partial}{\partial t} + \frac{1}{\hbar}\, \vb{\nabla}_{\vb{k}} \epsilon_e
	\cdot \vb{\nabla}_{\vb{r}}
	- \frac{e}{\hbar^2} \left( \vb{\nabla}_{\vb{k}} \epsilon_e \times \vb{B} \right)
	\cdot \vb{\nabla}_{\vb{k}} .
	\label{eq: diff op elec}
\end{equation}
For the coefficient of $\signull$, we obtain
\begin{equation}\label{eq:elec boltzmann n-e}
\mathcal{D}_e \meg - \frac{e}{\hbar}\, \vb{E} \cdot \vb{\nabla}_{\vb{k}} \meo
  = I^{n_e}[\depend]
\end{equation}
and for the coefficients of $\sigi$ with $i = 1,2,3$,
\begin{align}\label{eq:elec boltzmann c-ei}
	& \mathcal{D}_e \cegi - \frac{e}{\hbar} \left( \vb{E} +  \frac{1}{\hbar}\, \vb{\nabla}_{\vb{k}} \epsilon_e\times \vb{B} \right) \cdot \vb{\nabla}_{\vb{k}} \ceo B_i \nonumber\\
	&\quad{} - \frac{g_\text{el} \mu_B}{\hbar} \left(\vb{B} \times \ceg \right)_i
  = I^{c_e^i}[\depend] .
\end{align}

\subsection{Holes}

The description of holes in the valence band is analogous. Missing electrons close to the band top are described in terms of holes with positive mass and positive charge. The Hamiltonian for the holes reads as
\begin{equation}
\widehat{H}_h = \epsilon_h(\vb{k}) \signull - \frac{1}{2}\, g_\textrm{ho} \mu_B \vb{B} \cdot \sig ,
\end{equation}
where $g_\mathrm{ho}>0$ is the hole Land\'e factor.

The matrix-valued distribution function $\hole$ of the holes is decomposed into the equilibrium distribution and a deviation,
\begin{equation}
\hole = \holee + \holeg .
\end{equation}
The contributions are expanded according to
\begin{align}
\holee &= \frac{1}{2} \left( \mhe\,\signull + \che \vb{B} \cdot \sig \right) , \\
\holeg &= \frac{1}{2} \left( \mhg \signull + \chg \cdot \sig \right) .
\end{align}
The kinetic equations for the multipole coefficients $\mhg$ and $\chg$ can then be written down in analogy to the electronic case by replacing subscripts ``\textit{e}'' by ``\textit{h}'', $e$ by $-e$, and $g_\mathrm{el}$ by $-g_{\mathrm{ho}}$. Defining the differential operator
\begin{equation}
\mathcal{D}_h = \frac{\partial}{\partial t} + \frac{1}{\hbar}\, \vb{\nabla}_{\vb{k}} \epsilon_h
	\cdot \vb{\nabla}_{\vb{r}}
	+ \frac{e}{\hbar^2} \left( \vb{\nabla}_{\vb{k}} \epsilon_h \times \vb{B} \right)
	\cdot \vb{\nabla}_{\vb{k}} ,
\end{equation}
we obtain
\begin{align}
&\mathcal{D}_h \mhg + \frac{e}{\hbar}\, \vb{E} \cdot \vb{\nabla}_{\vb{k}} \mhe
  = I^{n_h}[\depend] , \\
&\mathcal{D}_h \chgi + \frac{e}{\hbar} \left( \vb{E}
  + \frac{1}{\hbar}\, \vb{\nabla}_{\vb{k}} \epsilon_h\times \vb{B} \right) \cdot \vb{\nabla}_{\vb{k}} \che B_i
    \nonumber\\
&\quad{} + \frac{g_\text{ho} \mu_B}{\hbar} \left(\vb{B} \times \chg \right)_i
  = I^{c_h^i}[\depend] .
\end{align}

\subsection{Singlet excitons}
\label{sub:singlet}

Singlet excitons are spin-$0$ quasiparticles. Consequently, the density $\singf$ and Hamiltonian $H_s$ are scalars so that the commutator term in Eq.\ (\ref{2.Boltzmann.3}) vanishes. This simplifies the Boltzmann equation to 
\begin{equation}
	\label{eq:singlet_boltzmann_general}
	\frac{\partial}{\partial t}\, \singf
	+ \frac{1}{\hbar}\, \vb{\nabla}_{\vb{k}} \epsilon_{s}  \cdot \vb{\nabla}_{\vb{r}} \singf
	= \scatsing{\depend} .
\end{equation}
Again, we use the group velocity with regard to the dispersion relation of the singlet excitons, $\epsilon_s(\vb{k})$. The description of excitons by a band structure $\epsilon_s(\vb{k})$ and an associated velocity has a long history \cite{Wan37,Dre56,Kan75} and has also been applied to organic semiconductors \cite{Dav71,VrS03,NPT21}. The excitonic band structure in pentacene has been studied experimentally by Schuster \textit{et al.}\ \cite{SKB07}. One should note that the band width and velocity of excitons can be larger than the ones of the charge carriers since the motion of excitons can take place through pure energy (F\"orster) transfer, without tunneling of electrons.

Similarly to the linear-response theory for electrons, we decompose the distribution function into an equilibrium distribution $\singfo$ and a deviation $\singg$,
\begin{equation}\label{eq:singlets decomp}
	\singf = \singfo + \singg .
\end{equation}
However, here the deviation is not related to the electric field and is not necessarily small since we also consider situations with optically excited excitons far from equilibrium. We assume $\singfo$ to be uniform. In equilibrium, excitons in semiconductors follow the Bose-Einstein distribution
\begin{equation}\label{eq:Bose-Einstein general}
	f_0(\epsilon) =
	\frac{1}{e^{(\epsilon-\mu)/k_BT} - 1} ,
\end{equation}
where $\epsilon$ is the exciton energy and $\mu$ is their chemical potential \cite{MoM00}. Inserting Eq.\ (\ref{eq:singlets decomp}) into Eq.\ (\ref{eq:singlet_boltzmann_general}) gives
\begin{equation}\label{eq:singlets boltzmann homogeneous}
 	\frac{\partial}{\partial t}\, \singg
 	+ \frac{1}{\hbar}\, \vb{\nabla}_{\vb{k}} \epsilon_s  \cdot \vb{\nabla}_{\vb{r}} \singg
 	= \scatsing{\depend} .
\end{equation}
The scalar form of the Boltzmann equation for singlet excitons allows one to obtain the solutions analytically. This requires specific assumptions for the scattering term. In a simple case, we can neglect interactions with other particle species and use the relaxation-time approximation, $I_s[\singf] =  - R_s\, ( \singf - \singfo ) = - R_s\, \singg$, to describe the scattering. $R_s$ is a possibly momentum-de\-pen\-dent relaxation rate. This transforms the Boltzmann equation into a homogeneous partial differential equation for $\singg$,
\begin{equation}\label{eq:singlets boltzmann R homogeneous}
 	\frac{\partial}{\partial  t}\, \singg
 	+ \frac{1}{\hbar}\, \vb{\nabla}_{\vb{k}} \epsilon_s \cdot \vb{\nabla}_{\vb{r}} \singg
 	+ R_s \singg = 0 .
\end{equation}
It is of first order and the coefficients are constant for fixed $\vb{k}$. The analytical solution can be derived using Fourier transformation with respect to $\vb{r}$. The result is
\begin{equation}
	\singg = g_s^0\, e^{- R_s t} \int \frac{d^3q}{(2\pi)^3} \:
	\exp\left( i\vb{q} \cdot \left[ \vb{r}
	  - \frac{1}{\hbar}\, \vb{\nabla}_{\vb{k}} \epsilon_{s} t\right] \right) ,
\end{equation}
with the integration constant $g_s^0$. This equation expresses an exponential dampening of the deviation that is superimposed onto a ballistic divergence in position space. This dynamics agrees with the expected behavior. We can take this as a confirmation of the form of the Boltzmann equation derived above.

\subsection{Triplet excitons}

Triplet excitons are spin-$1$ quasiparticles. Hence, their distribution function $\triplet$ is a $3\times 3$ matrix acting on the spin-$1$ Hilbert space and their Boltzmann equation takes the form
\begin{align}\label{eq:boltzmann f_T general}
	&\frac{\partial}{\partial t}\, \triplet + \frac{1}{\hbar}\, \vb{\nabla}_{\vb{k}} \epsilon_t
	  \cdot \vb{\nabla}_{\vb{r}} \triplet
	  + \frac{i}{\hbar}\, \big[\mspin{H}_t, \triplet\big]
	= \scatpart{\depend}{t} ,
\end{align}
with the triplet-exciton dispersion $\epsilon_t(\vb{k})$. As before, we decompose the distribution function into an equilibrium function and a deviation, $\triplet = \tripletfo + \tripletg$, and assume $\tripletfo$ to be uniform. Inserting this into Eq.\ (\ref{eq:boltzmann f_T general}) yields 
\begin{align} \label{eq:boltzmann f_T= f0,g}
	&\frac{\partial}{\partial t}\, \tripletg
	+ \frac{1}{\hbar}\, \vb{\nabla}_{\vb{k}} \epsilon_t
	\cdot \vb{\nabla}_{\vb{r}} \tripletg + \frac{i}{\hbar}\, \big[\mspin{H}_t, \tripletfo\big]
	+ \frac{i}{\hbar}\, \big[\mspin{H}_t, \tripletg\big] \nonumber \\
	&\quad = \scatpart{\depend}{t} ,
\end{align}
where the equilibrium commutator $[\mspin{H}_t, \tripletfo]$ again vanishes. We choose a basis of the space of Hermitian $3\times 3$ matrices in a similar fashion as for the $2\times 2$ case. This includes
\begin{equation}
\Snull = \sqrt{\frac{2}{3}}\: \one_3
\end{equation}
as the monopole term and the three-dimensional representation of $\mathrm{SU}(2)$,
\begin{align}
	\Sx &= \frac{1}{\sqrt{2}} \begin{pmatrix}
		0 & 1 & 0 \\
		1 & 0 & 1 \\
		0 & 1 & 0
	\end{pmatrix} , \\
	\Sy &= \frac{1}{\sqrt{2}} \begin{pmatrix}
		0 & -i & 0 \\
		i & 0 & -i \\
		0 & i & 0
	\end{pmatrix} , \\
	\Sz &= \begin{pmatrix}
		1 & 0 & 0 \\
		0 & 0 & 0 \\
		0 & 0 & -1
	\end{pmatrix} ,
\end{align}
as dipole terms. The normalization condition for all basis matrices is $\Tr \widehat{S}_i^2 = 2$. The remaining five matrices can be interpreted as quadrupole terms and are arranged as a five-component vector $\vec{\Q}$. We use the Hermitian matrices~\cite{AlH01}
\begin{align}\label{eq:q1-5 matrices}
	\Qx &= \big\{\Sy , \Sz\big\} = \frac{1}{\sqrt{2}} \begin{pmatrix}
		0 & 1 & 0 \\
		1 & 0 & -1 \\
		0 & -1 & 0
	\end{pmatrix} , \\
	\Qy &= \big\{\Sz , \Sx\big\} = \frac{1}{\sqrt{2}} \begin{pmatrix}
		0 & -i & 0 \\
		i & 0 & i \\
		0 & -i & 0
	\end{pmatrix} , \\
	\Qz &= \big\{\Sx , \Sy\big\} = \begin{pmatrix}
		0 & 0 & -i \\
		0 & 0 & 0 \\
		i & 0 & 0
	\end{pmatrix} , \\ 
	\Qxy &= \Sx^2 - \Sy^2 = \begin{pmatrix}
		0 & 0 & 1 \\
		0 & 0 & 0 \\
		1 & 0 & 0
	\end{pmatrix} , \\	
	\Qzr &= \sqrt{3}\, \big(\Sz^2 - \Snull^2\big) = \frac{1}{\sqrt{3}} \begin{pmatrix}
		1 & 0 & 0 \\
		0 & -2 & 0 \\
		0 & 0 & 1
	\end{pmatrix} ,
\end{align}
which are also normalized such that $\Tr \widehat{Q}_i^2 = 2$. The matrices $\Qx$, $\Qy$, and $\Qz$ transform as $t_{2g}$ under the cubic point group $O_h$, while $\Qxy$ and $\Qzr$ correspond to $e_g$.

The multipole expansion of $\triplet$ is
\begin{equation} \label{eq:f_t Q basis}
	\triplet = \mto \Snull + \ct \cdot \Sp + \qtv \cdot \vec{\Q} ,
\end{equation} 
with the coefficients $\mto \in \mathbb{R}$, $\ct \in \mathbb{R}^3$, $\qtv \in \mathbb{R}^5$. The related physical quantities are the number density $\sqrt{6}\, \mto$, the spin density $\ct$, and the quadrupole density $\qtv$. This is a good place to also introduce the Cartesian components of the quadrupole tensor operator,
\begin{equation}
\widehat{G}_{ij} = \big\{ \widehat{S}_i, \widehat{S}_j \big\}
  - \frac{4}{3}\, \delta_{ij} \one_3 ,
\end{equation}
which form the symmetric and traceless matrix
\begin{equation}
\gmat = \begin{pmatrix}
		\mspin{Q}_4 - \frac{1}{\sqrt{3}}\, \mspin{Q}_5 & \mspin{Q}_3 & \mspin{Q}_2 \\
		\mspin{Q}_3 & -\mspin{Q}_4 - \frac{1}{\sqrt{3}}\, \mspin{Q}_5 & \mspin{Q}_1 \\
		\mspin{Q}_2 & \mspin{Q}_1 & \frac{2}{\sqrt{3}}\, \mspin{Q}_5
	\end{pmatrix} .
\label{eq:def_gmat}
\end{equation}
Importantly, this matrix acts on the space of Cartesian spin components and each component is a $3\times 3$ matrix on the Hilbert space of a spin of length $1$. Using the Cartesian form, the quadrupolar term in the distribution function $\triplet$ can be rewritten as
\begin{equation}
\qtv \cdot \vec{\Q}
  = \frac{1}{2} \Tr_\mathrm{Cart} \matq\, \gmat ,
\end{equation}
where the trace is over Cartesian components and
\begin{equation}
\matq = \begin{pmatrix}
    \qtxy - \frac{1}{\sqrt{3}}\, \qtzr & \qtz & \qty \\
    \qtz &  - \qtxy - \frac{1}{\sqrt{3}}\, \qtzr & \qtx \\
    \qty & \qtx & \frac{2}{\sqrt{3}}\, \qtzr
   \end{pmatrix} .
\label{eq:def_qmat}
\end{equation}

Next, we motivate the form of the Boltzmann equation for these physical quantities. The Hamiltonian for the triplet excitons reads as
\begin{equation}
	\mspin{H}_t = \epsilon_t \one_3 + g_\mathrm{tr} \mu_B \vb{B} \cdot \Sp ,
\end{equation}
where $\epsilon_t(\vb{k})$ is the dispersion in the absence of a magnetic field (see the comments on excitonic band structures in Sec.\ \ref{sub:singlet}) and $g_\mathrm{tr}$ is the Land\'e factor of the triplet excitons, which is assumed to be scalar. A triplet exciton is of course charge neutral but since it generically consists of an electron and a hole with different Land\'e factors its total magnetic moment is nonzero. Additional terms linear in quadrupole matrices $\Q_i$ have been proposed in the literature \cite{Mer71,BaW16,Sal19}. This magneto-crystalline anisotropy is due to spin-orbit coupling and is therefore weak in organic semiconductors. We omit these terms for simplicity but implementing them in the Hamiltonian is straightforward.

If the external magnetic field is aligned along the $z$ axis, the Zeeman term simplifies to $g_\mathrm{tr} \mu_B B \Sz$ with $B = \left|\vb{B}\right|$. The Hamiltonian is diagonal in this case. We use this specific situation to derive the form of the equilibrium distribution function $\tripletfo$. The equilibrium distribution is the Bose-Einstein distribution and can immediately be evaluated for the diagonal components. The resulting $\tripletfo$ has a diagonal matrix structure as well,
\begin{equation} \label{eq:fT0 Q-ansatz Bz}
	\tripletfo = \mtoo \Snull + \ctzn B \Sz .
\end{equation}
A rotation yields the general description for an arbitrary direction of $\vb{B}$. The matrix $\Sz$ generalizes to $\Sp$ projected onto the direction of $\vb{B}$. Thus, the general form of the equilibrium distribution function is given by
\begin{equation} \label{eq:fT0 Q-ansatz general}
	\tripletfo = \mtoo \Snull + \cto \vb{B}\cdot \Sp .
\end{equation}
The deviation $\tripletg$ is of the most general form of the multipole expansion,
\begin{equation} \label{eq:gt Q-ansatz}
	\tripletg = \mtog \Snull + \ctg \cdot \Sp + \qtvg \cdot \vec{\Q} .
\end{equation}
The velocity is given by a derivative of the Hamiltonian of triplet excitons in analogy to Eq.\ (\ref{eq:evelocity_matrix}) and thus is a $3\times 3$ matrix. Since we assume spin-orbit coupling to be negligible, we can approximate it by a scalar, $\vb{v} = ({1}/{\hbar}) \vb{\nabla}_{\vb{k}} \epsilon_t$.

Kinetic equations for the coefficients can be obtained but the derivation is made complicated by  quadrupolar terms. The details are relegated to Appendix \ref{app.kinetic.triplet}. Defining the differential operator
\begin{equation} \label{eq:def Dt}
	\mathcal{D}_t = \frac{\partial}{\partial t}
	  + \frac{1}{\hbar}\, \vb{\nabla}_{\vb{k}} \epsilon_t \cdot \vb{\nabla}_{\vb{r}}
\end{equation}
and expanding the scattering integral into multipoles, we find the coupled equations
\begin{align}
	\mathcal{D}_t \mtog  &= I^{n_t}[\depend] ,
	\label{eq:boltzmann triplet 0} \\
  	\mathcal{D}_t \ctxg  &= \frac{g_\mathrm{tr} \mu_B}{\hbar} \,
  	  \big( B_2\ctzg - B_3\ctyg \big) + I^{c_t^x}[\depend] , \\
  	\mathcal{D}_t \ctyg  &= \frac{g_\mathrm{tr} \mu_B}{\hbar} \,
  	  \big( B_3\ctxg - B_1\ctzg \big) + I^{c_t^y}[\depend] , \\
  	\mathcal{D}_t \ctzg  &= \frac{g_\mathrm{tr} \mu_B}{\hbar}\,
  	  \big( B_1\ctyg - B_2\ctxg \big) + I^{c_t^z}[\depend] , \\
  	\mathcal{D}_t \qtxg  &= 
 	\frac{g_\mathrm{tr} \mu_B}{\hbar}\, \vb{B} 
		\cdot \begin{pmatrix}
 			-2 \qtxyg - 2\sqrt{3}\, \qtzrg  \\
 			- \qtzg   \\
			\qtyg  \\
 		\end{pmatrix} \nonumber \\
 	&\quad{} + I^{q_{t1}}[\depend] , \\
 	\mathcal{D}_t \qtyg  &= 
 	\frac{g_\mathrm{tr} \mu_B}{\hbar}\, \vb{B} 
		\cdot \begin{pmatrix}
 			\qtzg  \\
 			-2 \qtxyg + 2\sqrt{3}\, \qtzrg  \\
			- \qtxg  \\
 		\end{pmatrix} \nonumber \\
 	&\quad{} + I^{q_{t2}}[\depend] , \\
 	\mathcal{D}_t \qtzg  &=
		\frac{g_\mathrm{tr} \mu_B}{\hbar}\, \vb{B} 
		\cdot \begin{pmatrix}
 			- \qtyg  \\
 			\qtzg   \\
			4 \qtxyg \\
 		\end{pmatrix}
 	+ I^{q_{t3}}[\depend] , \\
 	\mathcal{D}_t \qtxyg  &= \frac{g_\mathrm{tr} \mu_B}{\hbar}\, \vb{B} 
		\cdot \begin{pmatrix}
 			\qtxg  \\
 			\qtyg   \\
			-2\qtzg \\
 		\end{pmatrix} + I^{q_{t4}}[\depend] , \\
  	\mathcal{D}_t \qtzrg  &= - \frac{g_\mathrm{tr} \mu_B}{\sqrt{3}\hbar}\, \vb{B} 
		\cdot \begin{pmatrix}
 			- 3\qtxg  \\
 			3\qtyg   \\
			0 \\
 		\end{pmatrix} + I^{q_{t5}}[\depend] .  \!
 	\label{eq:boltzmann triplet 8}
\end{align}
This complicated system cannot be simplified without further knowledge about the scattering terms. A detailed investigation of these integrals is presented in the following section.

\section{Spin-conserving scattering integrals}
\label{sec.scattering}

\begin{table*}
\caption{\label{tab.processes}Overview of scattering and transition processes treated explicitly in this paper.}
\begin{ruledtabular}
\begin{tabular}{lll}
Process & Description & Reference \\ \hline
$e\to e$ & disorder scattering of electrons & Sec.\ \ref{sec.scattering} (introduction) \\
$s\to s$ & disorder scattering of singlet excitons & Sec.\ \ref{sec.scattering} (introduction) \\
$|0\rangle \to s$ & generation of singlet exciton & Sec.\ \ref{sub.3A} \\
$s \to |0\rangle$ & decay of singlet exciton & Sec.\ \ref{sub.3A} \\
$e+h \to s$ & binding of electron and hole in singlet state & Sec.\ \ref{sub.3B} \\
$s \to e+h$ & unbinding of electron and hole in singlet state & Sec.\ \ref{sub.3B} \\
$e+h \to t$ & binding of electron and hole in triplet state & Sec.\ \ref{sub.3C} \\
$t \to e+h$ & unbinding of electron and hole in triplet state & Sec.\ \ref{sub.3C} \\
$2s \to s$ & fusion of singlet excitons & Sec.\ \ref{sub.3D} \\
$s \to 2s$ & fission of singlet exciton into singlet excitons & Sec.\ \ref{sub.3D} \\
$t+s \to t$ & absorption of singlet exciton by triplet exciton & Sec.\ \ref{sub.3E} \\
$2t\to s$ & fusion of triplet excitons into singlet exciton & Sec.\ \ref{sub.3F} \\
$s\to 2t$ & fission of singlet exciton into triplet excitons & Sec.\ \ref{sub.3F} \\
$e+t \to e$ & absorption of triplet exciton by electron & Sec.\ \ref{sub.3G} \\
$2t \to t$ & fusion of triplet excitons into triplet exciton & Sec.\ \ref{sub.3H} \\
\end{tabular}
\end{ruledtabular}
\end{table*}

In this section, we construct spin-conserving scattering integrals for the Boltzmann equations for electrons, holes, and excitons. These scattering integrals should not only describe collisions between quasiparticles and scattering off disorder but, importantly, also generation, decay, and transitions between excitons. An overview of the processes studied here is given in Table \ref{tab.processes}. The selection and order of these processes as well as the presentation in this section are partly pedagogical. We progress from conceptually simpler to more complicated cases, where essentially every subsection introduces an additional aspect, usually having to do with conservation of spin or indistinguishability of quasiparticles.
Additional processes can be treated in an analogous manner, e.g., the reverse of some of the processes in Table \ref{tab.processes} and processes for holes instead of electrons. Which processes are important of course depends on specific materials. Disorder scattering is expected to be always present. Bipolar semiconductors in which the carrier energies are sufficiently high in comparison to the exciton binding energies will also show electron-hole binding and unbinding, $e+h \leftrightarrow s$ or $e+h \leftrightarrow t$. For systems with photogenerated singlet excitons, e.g., photovoltaic devices, the fission $s\to 2t$ and the fusion $2t\to s$ are crucial, as noted in Sec.\ \ref{sec.introduction}. Note that the derivations in this section do not require any quasiparticle species to be close to equilibrium.

Scattering off disorder is described by collision integrals that are linear in the distribution function \cite{Zim72,AsM76}. Possanner and Negulescu \cite{PoN11} have studied such terms for both spin-conserving and spin-flip scattering, from a mathematical point of view. El Hajj \cite{Haj14} has studied spin-diffusion models derived from the resulting linear matrix Boltzmann equation in detail.

Our main interest is in the description of transitions between quasiparticle species. The required collision integrals are necessarily nonlinear. Quantum statistics imply that the transition rates depend on the occupation of the final states, contrary to what is assumed in the stochastic Liouville equation. For fermions, complete occupation of the final state prevents a process due to the Pauli principle, whereas for bosons, a large occupation of the final state enhances the rate. In the context of excitonic processes, this has been studied by Bisquert \cite{Bis08}, albeit not using an $\mathrm{SU}(2)$-invariant formalism.

It is useful to first outline the principles behind the construction of the collision integrals. Generation, decay, and transitions between excitons involve quasiparticles on different energy levels. This is reminiscent of laser theory, in which transition rates are derived by considering the occupation of the initial and final states. However, unlike in laser theory, these transition rates depend on momentum. Thus, they take a form similar to the basic scattering term in Boltzmann theory,
\begin{align}
I(\vb{k}) &= \int \frac{d^3k'}{(2\pi)^3}\, W(\vb{k}',\vb{k})\,
  f(\vb{k}')\, [ 1 - f(\vb{k}) ] \nonumber \\
&\quad{} - \int \frac{d^3k'}{(2\pi)^3}\, W(\vb{k},\vb{k}')\,
  f(\vb{k})\, [ 1 - f(\vb{k}') ] ,
\label{eq:Igeneral}
\end{align}
where $f$ is a distribution function. The two integrals describe in-scattering and out-scattering, respectively. The processes considered here are more complicated in a number of ways. First, our Boltzmann equations contain matrix-valued density functions. The idea is to derive the matrix form from the special case of \textit{z}-polarized particles by means of a rotation into a general direction, like discussed for the kinetic term in Sec.~\ref{sec.kineticeq}.

Second, the initial and final states can consist of more than one particle. We assume the initial state $A$ to contain particles of species $a_1,a_2,\ldots$ (some of which may be the same) and the final state $B$ to contain particles of species $b_1,b_2,\ldots$ (some of which may be the same). The integrand of the scattering integral is then proportional to
\begin{equation}
N_{a_1} N_{a_2} \cdots (1 \pm N_{b_1})(1 \pm N_{b_2}) \cdots,
\end{equation}
where $N_{a_1}$ etc.\ are the occupation numbers of states and the sign in factors $1\pm N_b$ is $+$ ($-$) if $b$ is bosonic (fermionic). Considering the limit $N_b \to 0$ yields the classical part of the process. For low concentrations of quasiparticles in the final state, in particular low exciton densities, this case is relevant  for applications.

In the following, we will progress from simple to more complicated cases. To fix the notation, we first discuss the essentially trivial case of elastic scattering of electrons off nonmagnetic impurities. It will be beneficial to write the matrix-valued distribution function of the electrons in components,
\begin{equation}
\label{eq:electron_densitymat}
\widehat f_e = \begin{pmatrix}
f_{e \uparrow \uparrow} & f_{e \uparrow \downarrow} \\
f_{e \downarrow \uparrow} & f_{e \downarrow \downarrow}
\end{pmatrix} .
\end{equation}
Since nonmagnetic scattering does not flip the spin, the matrix-valued collision integral appearing in the Boltzmann equation for $\widehat f_e$ is diagonal,
\begin{equation}
\widehat{I}^{\widehat{f}_e}_{e\to e} = \begin{pmatrix}
    I^{e\uparrow}_{e\to e} & 0 \\
    0 & I^{e\downarrow}_{e\to e}
  \end{pmatrix} .
\end{equation}
The collision integral for spin-up electrons can be written as~\cite{Zim72,AsM76}
\begin{align}
I^{e\uparrow}_{e\to e}(\vb{k}) &= \int \frac{d^3k'}{(2\pi)^3}\, W_{e'\to e}(\vb{k}',\vb{k})\,
  f_{e\uparrow}(\vb{k}')\, [ 1 - f_{e\uparrow}(\vb{k}) ] \nonumber \\
&\quad{} - \int \frac{d^3k'}{(2\pi)^3}\, W_{e\to e'}(\vb{k},\vb{k}')\,
  f_{e\uparrow}(\vb{k})\, [ 1 - f_{e\uparrow}(\vb{k}') ]
\label{eq:eup_disorder}
\end{align}
and analogously for spin-down electrons. Detailed balance for elastic scattering implies that $W_{e'\to e}(\vb{k}',\vb{k}) = W_{e\to e'}(\vb{k},\vb{k}')$ and thus
\begin{equation}
I^{e\uparrow}_{e\to e}(\vb{k}) = \int \frac{d^3k'}{(2\pi)^3}\, W_{e\to e'}(\vb{k},\vb{k}')\,
  [ f_{e\uparrow}(\vb{k}') - f_{e\uparrow}(\vb{k}) ]
\end{equation}
and analogously for spin down.

In the following, we employ a short-hand notation that suppresses (a) integrals over momenta $\vb{k}'$, $\vb{k}''$, \dots{} and (b) momentum arguments. Functions depending on $\vb{k}$, $\vb{k}'$, and $\vb{k}''$ are denoted by no, one, and two primes, respectively. Hence, we write the collision integrals for elastic impurity scattering as
\begin{align}
I^{e\uparrow}_{e\to e} &= W_{e\to e'}\, ( f_{e\uparrow}' - f_{e\uparrow} ) , \\
I^{e\downarrow}_{e\to e} &= W_{e\to e'}\, ( f_{e\downarrow}' - f_{e\downarrow} ) .
\end{align}
From Eq.\ (\ref{eq:multipole scattering term}), we then obtain the collision integrals for the electron number density,
\begin{equation}
I^{n_e}_{e\to e} = I^{e\uparrow}_{e\to e} + I^{e\downarrow}_{e\to e} = W_{e\to e'}\, ( n_e' - n_e ) ,
\label{eq:Ine_Iu_Id}
\end{equation}
and for the density of the \textit{z}-component of spin,
\begin{equation}
I^{c_e^z}_{e\to e} = I^{e\uparrow}_{e\to e} - I^{e\downarrow}_{e\to e} = W_{e\to e'}\, ( \ceza - c_e^z ) .
\label{eq:Icez_Iu_Id}
\end{equation}
The crucial next step is to reconstruct the $\mathrm{SU}(2)$-in\-va\-ri\-ant form of the collision integral for the spin by rotating all quantities into an arbitrary direction. In the present case, this is simple. We recognize that the left-hand side and both summands on the right-hand side are \textit{z}-components of vectors. The $\mathrm{SU}(2)$-invariant form is thus
\begin{equation}
\vb{I}^{\mathbf{c}_e}_{e\to e} = W_{e\to e'}\, ( \vb{c}_e' - \vb{c}_e ) .
\end{equation}
We can now use the multipole expansions in Eqs.\ (\ref{2a.f.multipole.3}) and (\ref{eq:multipole scattering term}) to express the collision rate in terms of the matrix densities as
\begin{equation}
\widehat I^{\widehat f_e}_{e\to e} = W_{e\to e'}\, ( \widehat f_e' - \widehat f_e ) .
\end{equation}
The case of holes is of course analogous.

The result for excitons is also analogous: For example for singlet excitons, the collision integral is
\begin{align}
&I^s_{s\to s}(\vb{k}) = \int \frac{d^3k'}{(2\pi)^3}\, W_{s'\to s}(\vb{k}',\vb{k})\,
  f_s(\vb{k}')\, [ 1 + f_s(\vb{k}) ] \nonumber \\
&\qquad{} - \int \frac{d^3k'}{(2\pi)^3}\, W_{s\to s'}(\vb{k},\vb{k}')\,
  f_s(\vb{k})\, [ 1 + f_s(\vb{k}') ] ,
\end{align}
where we have used that excitons are bosons. The second-order terms again cancel and we obtain
\begin{equation}
I^s_{s\to s} = W_{s\to s'}\, ( f_s' - f_s ) .
\label{eq.P2}
\end{equation}

\subsection{Generation and decay of singlet excitons}
\label{sub.3A}

Singlet excitons are described by the scalar distribution function $\gs$, so $\mathrm{SU}(2)$ invariance is automatically maintained. The transition rate describing the generation of singlet excitons only depends on the occupation of the final state. Since singlet excitons are bosons the resulting transition rate reads as
\begin{equation}
\rate{\ket{0} \rightarrow s}{s} = \widetilde{W}_{\ket{0} \rightarrow s}  \bracketr{1+ \gs} .
\end{equation}
Here $\ket{0} \rightarrow s$ is a short-hand way to denote that the singlet exciton is excited starting from the ground state with all single-electron states below (above) the Fermi energy occupied (empty). This is of course not a spontaneously occurring process but requires coupling to the radiation field, which is suppressed in this notation. We use the symbol $\widetilde{W}_{\ket{0} \rightarrow s}$ with a tilde for the rate; this is an effective rate with properties of the radiation field such as occupation numbers of photonic states absorbed. For the decay of a singlet exciton into the Fermi sea, we analogously write
\begin{equation}
\rate{s \rightarrow \ket{0}}{s} = - \widetilde{W}_{s \rightarrow \ket{0}} \gs .
\label{eq.P1}
\end{equation}
These transition rates can be added to the right-hand side of the Boltzmann equation\ (\ref{eq:singlet_boltzmann_general}) for the singlet excitons. The corresponding processes for triplet excitons are of course described analogously.

\subsection{Binding and unbinding of electrons and holes in a singlet state}
\label{sub.3B}

The transition between unbound electron-hole pairs and singlet excitons involves three different particle species, which results in a description from three different perspectives. We discuss these in turn.

\subsubsection{Point of view of the singlet exciton}

In the case of \textit{z}-polarized particles, there are only two different orientations of the electron and hole spins. To form a singlet exciton, a spin-up electron has to be paired with a spin-down hole or vice versa. Hence, the transition rate takes the form
\begin{equation}
\rate{e + h \rightarrow s}{s} = W_{e + h \rightarrow s}\, \frac{\feupa \fhdownd + \fedowna \fhupd}{2}\,
  (1 + \gs) .
\label{eq:eh_s_z_polarized}
\end{equation} 
Here functions depending on the momenta $\vb{k}'' = \vb{k} + \vb{k}'$ and $\vb{k}''' = \vb{k} - \vb{k}'$ are described by two and three primes, respectively. Note that the momentum arguments of the electron and the hole distribution functions could be interchanged as all possible combinations are covered by integrating over $\vb{k}'$. The origin of the factor $1/2$ is that the electron-hole states $\ket{\uparrow\downarrow}$ and $\ket{\downarrow\uparrow}$ can be written as $\ket{\uparrow\downarrow} = (\ket{\psi_s} + \ket{\psi_{t0}})/\sqrt{2}$ and $\ket{\downarrow\uparrow} = (-\ket{\psi_s} + \ket{\psi_{t0}})/\sqrt{2}$, respectively, where $\ket{\psi_s} = (\ket{\uparrow\downarrow}-\ket{\downarrow\uparrow})/\sqrt{2}$ is the singlet state and $\ket{\psi_{t0}} = (\ket{\uparrow\downarrow}+\ket{\downarrow\uparrow})/\sqrt{2}$ is the $m=0$ triplet state. Hence, the probability of, say, $\ket{\uparrow\downarrow}$ being in the singlet state is $1/2$.

For a spin along the $z$ axis, only the diagonal elements of the distribution function $\widehat f_e$ in Eq.\ (\ref{eq:electron_densitymat}) are nonzero and can be expressed in terms of each other. Rewriting Eq.\ (\ref{eq:eh_s_z_polarized}) in terms of the coefficients of the multipole expansion in Eq.\ (\ref{2a.f.multipole.3}) then reads as
\begin{equation}
\rate{e + h \rightarrow s}{s}
  = W_{e + h \rightarrow s}\, \frac{ \mea \mhd - \ceza \chzd}{4}\, (1 + \gs) .
\label{eq:eh_s_z_multipole_basis}
\end{equation}
Now it is possible to generalize this transition rate to arbitrary spin directions as there is a unique $\mathrm{SU}(2)$-in\-va\-ri\-ant expression that reduces to Eq.\ (\ref{eq:eh_s_z_multipole_basis}) for \textit{z}-polarized spins:
\begin{equation}
\rate{e + h \rightarrow s}{s}
= W_{e + h \rightarrow s}\, \frac{\mea \mhd - \cea \cdot \chd}{4}\, (1 + \gs) .
\label{eq:eh_s_multipole_basis}
\end{equation}
This transition rate $\rate{e + h \rightarrow s}{s} \equiv \rate{e+h \to s}{\gs}$ appears in the Boltzmann equation (\ref{eq:singlet_boltzmann_general}) for the scalar distribution function $\gs$ of singlet excitons and is thus itself a scalar. It can also be expressed in a basis-independent form in terms of matrix-valued distribution functions as~\cite{note.anticomm}
\begin{equation}
\rate{e+h \to s}{\gs} 
  = \frac{W_{e+h \to s}}{2}\, \Tr\anticomm{\psing, \electrona \otimes \holed}\, (1+\gs) ,
\label{eq:eh_s_matrixform}
\end{equation}
as can be shown by inserting the multipole expansions of the electron and hole distribution functions. Here $\{\bullet,\bullet\}$ is the anticommutator and $\psing$ is the projection operator onto the spin-singlet subspace of the product Hilbert space of the electron and hole spins. This operator can be written as
\begin{align}
\psing &= \frac{\one_4}{4} - \widehat{\vb{S}}_e \cdot \widehat{\vb{S}}_h \nonumber \\
&= \frac{1}{4} \left( \signull \otimes \signull
    - \sigx \otimes \sigx - \sigy \otimes \sigy - \sigz \otimes \sigz \right) ,
\label{eq:Ps_def}
\end{align}
where $\widehat{\vb{S}}_e$ and $\widehat{\vb{S}}_h$ are the spin operators of the electron and the hole, respectively. It expresses the fact that an electron and a hole can only form a singlet exciton if they are in a relative spin-singlet state. The structure of the momentum-dependent transition rate is reminiscent of the Haberkorn approach \cite{JoM70,EFL73,Hab76}. As mentioned in Sec.\ \ref{sec.introduction}, the anticommutator structure ensures positivity of the density matrix on the electron-hole product space. The trace in Eq.\ (\ref{eq:eh_s_matrixform}) sums over all contributions allowed by spin conservation and leads to a scalar transition term.

It is useful to restore the momentum arguments and integrals at this point. For the process $e+h\to s$ from the point of view of the exciton, the outer momentum $\vb{k}$ is the one of the exciton, the momentum $\vb{k}'$ of the electron needs to be integrated over and the momentum $\vb{k}''' = \vb{k} - \vb{k}'$ of the hole is then fixed by momentum conservation. This leads to
\begin{align}
\rate{e+h \to s}{\gs}(\vb{k}) &= \int \frac{d^3k'}{(2\pi)^3}\, \frac{W_{e+h \to s}(\vb{k}',\vb{k}-\vb{k}')}{2}
  \nonumber \\
&\quad{}\times
  \Tr\anticomm{\psing, \electron(\vb{k}') \otimes \hole(\vb{k}-\vb{k}')}\, \big[ 1+\gs(\vb{k}) \big] .
\end{align}

The breaking of singlet excitons into unbound electrons and holes can be described in the same way. For this reverse process only the sign and the occupation numbers have to be changed. Now the final state is fermionic, which yields
\begin{equation}
\rate{s \rightarrow e + h}{s}
  = - W_{s \rightarrow e + h}\,
  \frac{(2 - \mea)(2 - \mhd) - \cea \cdot \chd}{4}\, \gs
\end{equation}
in the multipole form and  
\begin{align}
\rate{s \rightarrow e + h}{\gs} &= - \frac{W_{s \to e+h}}{2} \nonumber \\
&\quad{}\times \Tr \Big\{ \psing, \big(\one_2  - \electrona\big)
  \otimes \big(\one_2 - \holed\big) \Big\}\, \gs
\end{align}
as the basis-independent expression; compare Eqs.\ (\ref{2a.f.multipole.3}) and~(\ref{eq:multipole scattering term}).

\subsubsection{Point of view of the electron}

We now consider the perspective of the electron (the hole case is of course analogous). The electron's momentum is the momentum argument $\vb{k}$ of the solution of the Boltzmann equation. The transition rates for spin-up and spin-down electrons read as
\begin{align}
\rate{e + h \rightarrow s}{e \uparrow} &= - W_{e + h \rightarrow s}\,
  \frac{\feup \fhdowna}{2}\, (1 + \gsc) , \\
\rate{e + h \rightarrow s}{e \downarrow} &= - W_{e + h \rightarrow s}\,
  \frac{\fedown \fhupa}{2}\, (1 + \gsc) ,
\end{align}
respectively. The transition rate for the electron number density is the sum of these two transition rates, while the transition rate of the density of the $z$-component of the spin is their difference. After generalizing to an arbitrary polarization direction, we obtain
\begin{equation}
\rate{e + h \rightarrow s}{\me} = - W_{e + h \rightarrow s}\,
  \frac{ \me \mha- \ce \cdot  \cha}{4}\, (1 + \gsc)
\end{equation}
and
\begin{equation}
\ratec{e + h \rightarrow s}{\ce} = - W_{e + h \rightarrow s}\,
  \frac{ \ce \mha - \cha \me}{4}\, (1 +\gsc).
\end{equation}
The transition rate of the electron number density is the negative of the transition rate from the point of view of the singlet exciton given in Eq.\ (\ref{eq:eh_s_multipole_basis}). According to Eqs.\ (\ref{2a.f.multipole.3}) and (\ref{eq:multipole scattering term}), the matrix-valued transition term then reads as
\begin{align}
\widehat I^{\widehat f_e}_{e+h\to s}
&= \frac{1}{2}\, \big(\rate{e + h \rightarrow s}{\me} \signull
  + \ratec{e + h \rightarrow s}{\ce} \cdot \sig\big) \nonumber \\
&= -\frac{ W_{e+h \to s} }{2}\, \Tr_h \big\{\psing, \electron \otimes \holea\big\}\, (1+ \gsc),
\label{eq:eh_to_s_matrixform}
\end{align}
where $\Tr_h$ is the trace over the hole sector. To restore the momenta and integrals, we note that the outer momentum $\vb{k}$ is the one of the electron, the hole momentum $\vb{k}'$ is integrated over, and the exciton momentum $\vb{k}'' = \vb{k} + \vb{k}'$ is then fixed. The rate reads as
\begin{align}
\widehat I^{\widehat f_e}_{e+h\to s}(\vb{k}) &= - \int \frac{d^3k'}{(2\pi)^3}\,
  \frac{W_{e+h \to s}(\vb{k},\vb{k}')}{2} \nonumber \\
& ~{}\times \Tr_h \big\{\psing, \electron(\vb{k}) \otimes \hole(\vb{k}')\big\}\,
  \big[ 1+ \gs(\vb{k} + \vb{k}') \big] .
\end{align}

These transition rates can be compared to the transition rates from the point of view of the singlet exciton. This yields
\begin{equation}
{-}\rate{e + h \rightarrow s}{s} = \rate{e + h \rightarrow s}{\me}= \rate{e + h \rightarrow s}{\mh},
\end{equation}
as the creation of one singlet exciton results in the annihilation of a free electron and a free hole. Furthermore, the spin transition rates satisfy
\begin{equation}
\ratec{e + h \rightarrow s}{\ce} + \ratec{e + h \rightarrow s}{\ch} = 0 ,
\end{equation}	
which shows that spin is conserved, as the singlet exciton does not posses an overall spin.

The reverse unbinding process can be described analogously. It can also be read off from Eq.\ (\ref{eq:eh_to_s_matrixform}), essentially by reversing the sign and interchanging occupied and empty states. The resulting rate reads as
\begin{equation}
\widehat I^{\widehat f_e}_{s \to e+h} = \frac{ W_{s\to e+h} }{2}\,
  \Tr_h \Big\{\psing, \big( \one_2 - \electron \big) \otimes \big( \one_2 - \holea \big) \Big\}\, \gsc .
\end{equation}

\subsection{Binding and unbinding of electrons and holes in a triplet state}
\label{sub.3C}

Like for electrons and holes, it is beneficial to write the matrix-valued distribution function of the triplet excitons in components,
\begin{equation}
\triplet = \begin{pmatrix}
  f_{t, 11} & f_{t, 10} & f_{t, 1\bar{1}} \\
  f_{t, 01} & f_{t, 00} & f_{t, 0\bar{1}} \\
  f_{t, \bar{1}1} & f_{t, \bar{1}0} & f_{t, \bar{1}\bar{1}}
\end{pmatrix}.
\end{equation}
The indices of the matrix elements denote the magnetic quantum number of the triplet exciton, where we use $\bar{1} = -1$. For a spin along the $z$ axis, only the diagonal elements of the matrix are nonzero so that in the multipole representation it takes the form
\begin{equation}
\triplet = \begin{pmatrix}
    f_{t,1} & 0 & 0 \\
    0 & f_{t,0} & 0 \\
    0 & 0 & f_{t,\bar{1}}
  \end{pmatrix}
  = \mto \Snull + \ctz \widehat{S}_3 + \qtzr \Q_5 ;
\end{equation}
cf.\ Eq.\ (\ref{eq:f_t Q basis}). This allows us to express the elements of the density matrix in terms of the diagonal elements of the multipole expansion and vice versa, 
\begin{align}
\mto &= \frac{f_{t,1} + f_{t,0} + f_{t,\bar{1}}}{\sqrt{6}} , \\
\ctz &= \frac{f_{t,1} - f_{t,\bar{1}}}{2} , \\
\qtzr &= \frac{f_{t,1} - 2f_{t,0} + f_{t,\bar{1}}}{2\sqrt{3}} .
\end{align}

\subsubsection{Point of view of the triplet exciton}

The transition term for the triplet-exciton distribution function $\widehat f_t$ also has to be a $3\times 3$ matrix on the spin-$1$ Hilbert space. For the spin along the $z$ axis, spin conservation implies that we only need to consider the transition rates of the diagonal components, which read as
\begin{align}
\rate{e+h \rightarrow t}{\tp} &= W_{e+h \rightarrow t}\, \feupa \fhupd\, (1+\ftp) , \\
\rate{e+h \rightarrow t}{\tz} &= W_{e+h \rightarrow t}\, \frac{\feupa \fhdownd + \fedowna \fhupd}{2}\,
  (1+\ftz) , \\
\rate{e+h \rightarrow t}{\tm} &= W_{e+h \rightarrow t}\, \fedowna \fhdownd\, (1+\ftm) .
\end{align}
Recall that three primes denote a dependence on $\vb{k}''' = \vb{k} - \vb{k}'$. The next step is again to construct the $\mathrm{SU}(2)$-invariant transition rates by rotating the spin into a general direction. In analogy to the multipole expansion of the distribution function in Eq.\ (\ref{eq:f_t Q basis}), we expand the transition term as
\begin{equation}
\widehat I^{\widehat f_t}_{e+h \rightarrow t}
  = I^{n_t}_{e+h \rightarrow t} \Snull + \vb{I}^{\ct}_{e+h \rightarrow t} \cdot \Sp
  + \vec I^{\,\qtv}_{e+h \rightarrow t} \cdot \vec{\Q} .
\end{equation}
For the spin along the $z$ direction, this yields
\begin{align}
\rate{e+h \to t}{\mto} &=
  \frac{\rate{e+h \rightarrow t}{\tp} + \rate{e+h \rightarrow t}{\tz}
  + \rate{e+h \rightarrow t}{\tm}}{\sqrt{6}} \nonumber \\
&= \frac{W_{e+h \rightarrow t}}{24}\, \big(
  3\sqrt{6}\, n_e' n_h''' + \sqrt{6}\, \ceza \chzd \nonumber \\
&\quad{}+ 6\, n_e' n_h''' n_t
  + 2\, \ceza \chzd n_t
  + 2\sqrt{6}\, \ceza n_h''' c_t^z \nonumber \\
&\quad{}+ 2\sqrt{6}\, n_e' \chzd c_t^z
  + 4\sqrt{2}\, \ceza \chzd q_{t5} \big)
\label{eq:I_nt_zpol}
\end{align}
and analogously
\begin{align}
\rate{e+h \to t}{\ctz} &=
  \frac{W_{e+h \rightarrow t}}{12}\, \big(
  3\, n_e' \chzd + 3\, \ceza n_h''' \nonumber \\
&\quad{}+ 3\, n_e' n_h''' c_t^z
  + \sqrt{6}\, n_e' \chzd n_t
  + \sqrt{6}\, \ceza n_h''' n_t \nonumber \\
&\quad{}+ 3\, \ceza \chzd c_t^z
  + \sqrt{3}\, n_e' \chzd q_{t5} + \sqrt{3}\, \ceza n_h''' q_{t5} \big) ,
\label{eq:I_ctz_zpol} \\
\rate{e+h \to t}{\qtzr} &=
  \frac{W_{e+h \rightarrow t}}{12\sqrt{3}}\, \big(
  6\, \ceza \chzd \nonumber \\
&\quad{}+ 3\sqrt{3}\, n_e' n_h''' q_{t5}
  + 3\, n_e' \chzd c_t^z + 3\, \ceza n_h''' c_t^z \nonumber \\
&\quad{}+ 2\sqrt{6}\, \ceza \chzd n_t
  - \sqrt{3}\, \ceza \chzd q_{t5} \big) .
\label{eq:I_qtzr_zpol}
\end{align}
In these equations, the terms of second order in the coefficients represent the classical part since they do not depend on the occupation of the final state. The terms of third order are the quantum corrections due to the Bose-Einstein statistics of the excitons.

Unlike for the previous examples, the $\mathrm{SU}(2)$-invariant generalization of Eqs.\ (\ref{eq:I_nt_zpol})--(\ref{eq:I_qtzr_zpol}) is not obvious since these transition rates contain the quadrupole coefficient $\qtzr$. The transformation of the coefficient vector $\vec{q}_t$ under rotations is not clear. On the other hand, the Cartesian quadrupole density $\matq$ defined in Eq.\ (\ref{eq:def_qmat}) transforms like a matrix under rotations. To make use of this, we have to identify the factors multiplying $\qtzr$ as components of Cartesian matrices and write Eqs.\ (\ref{eq:I_nt_zpol})--(\ref{eq:I_qtzr_zpol}) as components of proper matrix products.

This can be done by considering the two-particle state of an electron and a hole. As the unbound electron-hole pair can be of triplet character, it generally has a quadrupole moment. The
projection operator
\begin{equation}
\ptrip = \one_4 - \psing = \frac{3\one_4}{4} + \widehat{\vb{S}}_e \cdot \widehat{\vb{S}}_h
\end{equation}
projects onto the triplet subspace, in analogy to the singlet projection operator $\psing$ defined in Eq.\ (\ref{eq:Ps_def}). Since this subspace is three dimensional the projected density matrix $\ptrip \big( \electrona \otimes \holed \big) \ptrip$ can be written as a $3\times 3$ matrix with respect to a suitable basis of this subspace. To be consistent, we choose the canonical spin-$1$ basis $\{|{\uparrow_e\uparrow_h}\rangle, (|{\uparrow_e\downarrow_h}\rangle + |{\downarrow_e\uparrow_h}\rangle)/\sqrt{2}, |{\downarrow_e\downarrow_h}\rangle\}$. The notation $\bracket{\bullet}_3$ refers to the corresponding matrix. The density matrix for an electron-hole pair in a triplet state is thus written as
\begin{equation}
\label{eq:def.intermed}
\intermed = \Big[ \ptrip\, \big(\electrona \otimes \holed \big)\, \ptrip \Big]_3 .
\end{equation}
The resulting matrix $\intermed$ is an operator on the Hilbert space of spin-1 particles and can thus be expressed in terms of the nine basis matrices of the multipole expansion.
For a spin along the $z$ axis, only three coefficients are nonzero, $\intermed = n_{eh} \Snull + c_{eh}^z \widehat{S}_3 + q_{eh5} \Q_5$, where
\begin{align}
n_{eh} &= \frac{ \ceza \chzd + 3 \mea \mhd}{4 \sqrt{6}},
\label{eq:neh.eh} \\
c_{eh}^z &= \frac{\ceza \mhd + \chzd \mea}{4}, \\
q_{eh5} &=\frac{\sqrt{3}\, \ceza \chzd}{6}.
\label{eq:qeh.eh}
\end{align}
As electron and hole together have the overall momentum $\vb{k}$ due to momentum conservation, $\intermed$ and the corresponding multipole coefficients do not carry any primes. By eliminating the variables $\mea$, $\mhd$, $\ceza$, and $\chzd$ from Eqs.\ (\ref{eq:I_nt_zpol})--(\ref{eq:I_qtzr_zpol}) and (\ref{eq:neh.eh})--(\ref{eq:qeh.eh}), we obtain
\begin{align}
\rate{e+h \to t}{\mto} &= \frac{W_{e+h \rightarrow t}}{3}\, \big( 3 n_{eh} + \sqrt{6}\, n_{eh} n_t
  + \sqrt{6}\, c_{eh}^z c_t^z \nonumber \\
&\quad{} + \sqrt{6}\, q_{eh5} q_{t5} \big) , \\
\rate{e+h \to t}{\ctz} &=
  \frac{W_{e+h \rightarrow t}}{3}\, \big( 3 c_{eh}^z + \sqrt{6}\, n_{eh} c_t^z
  + \sqrt{3}\, q_{eh5} c_t^z \nonumber \\
&\quad{} + \sqrt{6}\, c_{eh}^z n_t + \sqrt{3}\, q_{t5} c_{eh}^z \big) , \\
\rate{e+h \to t}{\qtzr} &=
  \frac{W_{e+h \rightarrow t}}{3}\, \big( 3 q_{eh5} + \sqrt{6}\, q_{eh5} n_t + \sqrt{6}\, n_{eh} q_{t5}
  \nonumber \\
&\quad{} + \sqrt{3}\, c_{eh}^z c_t^z - \sqrt{3}\, q_{eh5} q_{t5} \big) .
\end{align}
These relations allow us to construct the $\mathrm{SU}(2)$-invariant generalizations without ambiguities. For terms containing quadrupole moments, Eq.\ (\ref{eq:def_qmat}) should be noted. The transition rate of the triplet exciton number density is
\begin{align}
\rate{e+h \rightarrow t}{\mto}
  &= \frac{W_{e+h \rightarrow t}}{6}\, \big( 6\, \mtoeh
  + 2\sqrt{6}\, \mtoeh \mto
  + 2\sqrt{6}\, \ceh \cdot \ct \nonumber \\
&\quad{}+ \sqrt{6}\, \Tr \matqeh \matq \big) ,
\end{align}
where the classical part is proportional to $\mtoeh$. For the last term, the identity $\Tr \matqeh \matq = 2\qeh \cdot \qt$ has been used. The spin transition rate becomes
\begin{align}
\ratec{e+h \rightarrow t}{\ct}
  &= \frac{W_{e+h \rightarrow t}}{6}\, \big( 6\, \ceh
  + 2\sqrt{6}\, \mtoeh \ct
  + 3\, \matqeh \ct \nonumber \\
&\quad{}+ 2\sqrt{6}\, \ceh \mto + 3\, \matq \ceh \big) ,
\end{align}
where the classical part is proportional to $\ceh$. The transition rate of the quadrupole part has to be traceless and symmetric in order to ensure that the full triplet-exciton transition rate is Hermitian. This requires the subtraction of the diagonal parts of matrices obtained by a naive restoration of $\mathrm{SU}(2)$ symmetry and yields
\begin{align}
\ratem{e+h \rightarrow t}{\matq}
&= W_{e+h \rightarrow t}\, \bigg[
	\matqeh + \frac{\sqrt{6}}{3}\, \matqeh \mto
	+ \frac{\sqrt{6}}{3}\, \mtoeh \matq \nonumber \\
&\quad{}+ \frac{1}{2}\, ( \ct \otimes \ceh + \ceh \otimes \ct )
  - \frac{1}{3}\, (\ceh \cdot \ct)\, \one_3 \nonumber \\
&\quad{}- \frac{1}{2}\, \big( \matqeh \matq  + \matqeh \matq \big)
  + \frac{1}{3}\, \big(\Tr \matqeh \matq\big)\, \one_3 \bigg] ,
\end{align}
with the classical part being proportional to $\matqeh$.

It is also useful to express the triplet-exciton transition rate in a basis-independent form. In contrast to the formation of a singlet exciton, the final state is now also described by a matrix-valued distribution function. The resulting transition rate takes the form
\begin{equation}
\widehat I^{\widehat f_t}_{e+h \rightarrow t}
  = \frac{W_{e+h \to t}}{2}\, \Big\{ \Big[ \ptrip\, \big(\electrona \otimes \holed\big)\, \ptrip \Big]_3,
    \one_3 + \triplet \Big\},
\end{equation}
where $\intermed$ as defined in Eq.\ (\ref{eq:def.intermed}) appears. For the reverse process $t\to e+h$, the matrix
\begin{equation}
\intermedinv = \ptrip\, \Big[ \big( \one_2 - \electrona \big) \otimes \big( \one_2 - \holed \big) \Big]\,
  \ptrip
\end{equation}
occurs instead, which results in a similar structure. In particular, the basis-independent transition rate from the point of view of the triplet excitons reads as
\begin{align}
\widehat I^{\widehat f_t}_{t \to e+h}
  &= - \frac{W_{t \to e+h}}{2} \nonumber \\
&\quad{} \times \Big\{ \Big[ \ptrip\, \Big[ \big( \one_2 - \electrona \big)
    \otimes \big( \one_2 - \holed \big) \Big]\, \ptrip \Big]_3, \triplet \Big\} ,
\end{align}
from which the rates for the number, spin, and quadrupole densities can be inferred.

\subsubsection{Point of view of electron and hole}
\label{subsub.eh_to_t.viewe}

In the following, we discuss the process $e+h \to t$ from the point of view of the electrons; the results for the holes are analogous. The transition rates of the different spin orientations for \textit{z}-polarized electrons,
\begin{align}
\rate{e+h \rightarrow t}{e \uparrow} &= -W_{e+h \rightarrow t}\, \bigg[ \feup \fhupa\, (1+\ftpc) \nonumber \\
&\quad{}+ \frac{\feup \fhdowna}{2}\, (1+\ftzc) \bigg] , \\
\rate{e+h \rightarrow t}{e \downarrow} &= -W_{e+h \rightarrow t}\, \bigg[ \fedown \fhdowna\, (1+\ftmc)
  \nonumber \\
&\quad{}+ \frac{\fedown \fhupa}{2}\, (1+\ftzc) \bigg] ,
\end{align}
can be combined to obtain the transition rates for the electron number and spin densities in analogy to Eqs.\ (\ref{eq:Ine_Iu_Id}) and (\ref{eq:Icez_Iu_Id}). The resulting expressions are not symmetric in the coefficients of $\electron$ and $\holea$ for the simple reason that we are writing down transition rates for the electrons, not for the holes. The rates are not sufficient to read off the $\mathrm{SU}(2)$-invariant form, as we will see shortly.

The basis-independent form of the transition rate is particularly useful for this reason. In analogy to the previous cases, the transition rate in terms of density matrices can be written as
\begin{align}
\widehat I_{e+h \to t}^{\electron} &= - \frac{W_{e+h\to t}}{2} \Tr_h \bigg[  \bracketr{\one_3 + \tripletc}_4
   \ptrip \bracketr{\electron \otimes \holea} \nonumber \\
&\quad{}+ \bracketr{\electron \otimes \holea} \ptrip
   \bracketr{\one_3 + \tripletc}_4 \bigg] .
\label{eq:ehtot.fullIfe}
\end{align}
The notation $\bracketr{\bullet}_4$ means that a $3\times 3$ matrix on the triplet subspace is extended to the full two-particle space by adding a $1\times 1$ null matrix on the singlet subspace. The matrix is then transformed into the product basis of electron and hole spins, $\{\ket{\uparrow_e}\ket{\uparrow_h}, \ket{\uparrow_e}\ket{\downarrow_h}, \ket{\downarrow_e}\ket{\uparrow_h}, \ket{\downarrow_e}\ket{\downarrow_h}\}$. The partial trace over the hole sector is defined by
\begin{equation}
\Tr_h \bullet = \bra{\uparrow_h} \bullet \ket{\uparrow_h} + \bra{\downarrow_h} \bullet \ket{\downarrow_h} .
\end{equation}
The resulting $\mathrm{SU}(2)$-invariant transition rates given in the multipole expansion are thus
\begin{align}
\rate{e+h \rightarrow t}{\me}
&= - \frac{ W_{e+h \rightarrow t}}{12}\, \big[ 9 \me \mha + 3 \ce \cdot\cha
  + 3 \sqrt{6}\, \me \mha \mtoc \nonumber \\
&\quad{} + 6 \me \cha \cdot \ctc + 6 \mha \ce \cdot \ctc + \sqrt{6}\, \mtoc \ce \cdot \cha \nonumber \\
&\quad{} + 6 \ce \cdot \matqc \cha \big]
\end{align}
for the electron number density and
\begin{align}
\ratec{e+h \rightarrow t}{\ce} &=- \frac{W_{e+h \rightarrow t}}{12}\,
  \big[ 9 \ce \mha + 3 \me \cha \nonumber \\
&\quad{} + 3 \sqrt{6}\, \ce \mha \mtoc + \sqrt{6}\, \me \cha \mtoc + 6 \me \mha \ctc \nonumber \\
&\quad{} + 6 \ce  \bracketr{\cha \cdot \ctc}
  + 6 \me \matqc \cha \big]
\end{align}
for the spin density. This rate contains a product of three spin densities. The corresponding term is proportional to $c_e^z c_h^{z\prime} c_t^{z\prime\prime}$ in the limit of \textit{z}-polarized spins. It is thus clear that this limit does not allow to infer the full $\mathrm{SU}(2)$-invariant expression (although one might guess correctly) so that the basis-independent expression (\ref{eq:ehtot.fullIfe}) is required.

The results for the holes are analogous. We thus find that the rates satisfy the consistency relations
\begin{align}
-\sqrt{6}\, \rate{e+h \to t}{\mto} &= \rate{e+h \to t}{\me} = \rate{e+h \to t}{\mh} , \\
-4\ratec{e+h \to t}{\ct}\, &= \ratec{e+h \to t}{\ce} + \ratec{e+h \to t}{\ch} .
\end{align}
Note that the factors $\sqrt{6}$ and $4$ in these two equations stem from the different normalization conditions concerning electrons and triplet excitons. The reverse process can be treated similarly.

\subsection{Fusion and fission of singlet excitons}
\label{sub.3D}

In preparation for the important transitions between a singlet exciton and two triplet excitons, $s\leftrightarrow 2t$, we first consider the simpler process $s\leftrightarrow 2s$. In contrast to the processes investigated before, the fusion and fission of singlet excitons involves the same particle species in the initial and final states. Furthermore, the two singlet excitons on the same side of the reaction have to be treated as indistinguishable bosons.

Starting with the fusion, this process can either create or fuse a singlet exciton with the outer momentum~$\vb{k}$. This leads to in-scattering and out-scattering terms in the transition rate, which can be written as
\begin{equation}
\rate{2s \to s}{\gs} =  W'''_{2s \to s}\gs''' \gs' \bracketr{1+ \gs}- 2W_{2s \to s} \gs \gs' \bracketr{1+\gs''}.
\end{equation}
Here the short-hand notations $W \equiv W(\vb{k}', \vb{k})$ and $W''' \equiv W(\vb{k}', \vb{k} - \vb{k}') \equiv W(\vb{k}', \vb{k}''')$ are used. The factor $2$ in the out-scattering term results from the two equal contributions of indistinguishable singlet excitons in the initial state. Conversely, the fission of one singlet exciton into two singlet excitons is described by
\begin{align}
\rate{s \to 2s}{\gs} &= 2W_{s \to 2s} \bracketr{1+ \gs} \bracketr{1+\gs'} \gs'' \notag \\
&\quad{} - W'''_{s \to 2s}\bracketr{1+ \gs'''} \bracketr{1+\gs'} \gs.
\end{align}
Note that due to the spin-singlet character of all involved quasiparticles, $\mathrm{SU}(2)$ symmetry is trivially satisfied.

\subsection{Absorption of a singlet exciton by a triplet exciton}
\label{sub.3E}

In contrast to the previous process, the particles in the initial state are distinguishable. In the limit of $z$ polarization, this yields, from the point of view of the singlet exciton,
\begin{align}
\rate{t+s \to t}{s} &= -W_{t+s \to t}\, \big[
  \ftpa (1+ \ftpc) + \ftza  (1+\ftzc) \nonumber \\
&\quad{}+ \ftma  (1+\ftmc) \big] \gs .
\end{align}
In terms of the multipole expansion this becomes
\begin{align}
\rate{t+s \to t}{s} &= -W_{t+s \to t}\, \big( \sqrt{6}\, \mtoa \nonumber \\
&\quad{} + 2 \mtoa \mtoc + 2 \cta \cdot \ctc + 2 \qta \cdot \qtc \big) \gs .
\end{align}
The singlet exciton is described by a scalar distribution function so that the basis-independent transition rate is
\begin{equation}
\rate{t+s \to t}{s}
  = - \frac{W_{t+s \to t}}{2}\, \Tr \Big\{\tripleta, \one_3 +  \tripletc\Big\}\, \gs.
\end{equation}

The transition term from the perspective of the triplet exciton features in-scattering and out-scattering terms as the triplet exciton is present on both sides of the reaction. The transition rates for the different spin orientations are
\begin{align}
\rate{t+s \to t}{\tp} &= W'''_{t+s \to t}\ftpa \gs'''\bracketr{1 + \ftp} \notag \\
&\quad{} - W_{t+s \to t} \ftp  \gs' \bracketr{1 + \ftpc},\\
\rate{t+s \to t}{\tz} &= W'''_{t+s \to t} \ftza \gs'''\bracketr{1 + \ftz} \notag \\
&\quad{} - W_{t+s \to t}\ftz  \gs' \bracketr{1 + \ftzc},\\
\rate{t+s \to t}{\tm} &= W'''_{t+s \to t} \ftma \gs'''\bracketr{1 + \ftm} \notag \\
&\quad{} - W_{t+s \to t}\ftm  \gs' \bracketr{1 + \ftmc}.
\end{align}
Since the (positive) in-scattering and the (negative) out-scattering contributions have the same structure it is sufficient to investigate one type. For the out-scattering contributions to the $\mathrm{SU}(2)$-invariant transition rates of the multipole coefficients we obtain
\begin{align}
\rate{t+s \to t}{\mto,\,\mathrm{out}} &= -\frac{W_{t+s \to t}}{\sqrt{6}}\, \big( \sqrt{6}\, \mtob
  \nonumber \\
&\quad{}+ 2 \mtob \mtoc + 2 \ctb \cdot \ctc + \Tr \matqb \matqc \big) \gs' , \\
\ratec{t+s \to t}{\ct,\,\mathrm{out}} &= -\frac{W_{t+s \to t}}{6}\, \big( 6 \ctb
  + 2\sqrt{6}\, \mtob \ctc + 2\sqrt{6}\, \ctb \mtoc \nonumber \\
&\quad{}+ 3 \matqb \ctc + 3 \matqc \ctb \big) \gs' , \\
\ratem{t+s \to t}{\matq,\,\mathrm{out}} &= -W_{t+s \to t}\, \bigg[ \matqb
  + \frac{\sqrt{6}}{3}\, \big( \mtob \matqc + \matqb \mtoc \big) \nonumber \\
&\quad{} + \frac{1}{2}\, \big(\ct \otimes \ctc + \ctc \otimes \ct\big)
  - \frac{1}{3}\, \ctc \cdot \ct \one_3 \nonumber \\
&\quad{} - \frac{1}{2}\, \big( \matqc \matqb + \matqb \matqc \big)
  + \frac{1}{6}\, \Tr {\matq \matqc} \one_3 \bigg] \gs' ,
\end{align}
where the quadrupole transition rate is again symmetric and traceless. In a basis-independent form, these relations can be summarized as
\begin{align}
\rate{t+s \to t}{\triplet} &= \frac{W'''_{t+s \to t}}{2} \big\{\tripleta, \one_3 + \tripletb \big\} \gs'''
  \notag \\
&\quad{} - \frac{W_{t+s \to t}}{2} \big\{\tripletb, \one_3 + \tripletc \big\} \gs' .
\end{align}
The treatment of singlet-exciton emission is analogous.

\subsection{Fusion and fission involving two triplet excitons and one singlet exciton}
\label{sub.3F}

The important fusion of two triplet excitons forming a singlet exciton \cite{StU89} is similar to the fusion of electron and hole into a singlet exciton. In contrast to the latter process, the two particles forming the singlet exciton are indistinguishable bosons.

\subsubsection{Point of view of the singlet exciton}

Considering the Clebsch-Gordan coefficients, the transition rate in terms of diagonal elements of the distribution function is
\begin{equation}
\rate{2t \to s}{s} = \frac{W'''_{2t \to s}}{3}\, \big( \ftpa \ftmd + \ftza \ftzd + \ftma \ftpd \big)
  (1+\gs) .
\end{equation}
Written in the multipole expansion and generalized to the $\mathrm{SU}(2)$-invariant form this yields
\begin{equation}
\rate{2t \to s}{s} = \frac{2W'''_{2t \to s}}{3}\, \big( \mtoa \mtod - \cta \cdot \ctd
  + \qta \cdot \qtd \big) (1 + \gs) ,
\label{eq:singlet_fission_singlet}
\end{equation}
where the relation $\Tr \matqa \matqd = 2\qta \cdot \qtd$ has been used. The only differences compared to the creation of singlet excitons from free charge carriers are the prefactor and the additional quadrupole term. The ba\-sis-in\-de\-pen\-dent form is
\begin{equation}
\rate{2t \to s}{\gs}
  = \frac{W'''_{2t \to s}}{2}\, \Tr \big\{ \psing, \tripleta \otimes \tripletd \big\} \bracketr{1+\gs} ,
\label{eq:singlet_fission_singlet_basind}
\end{equation}
which is analogous to the basis-independent form in the process $e+h \to s$; see Eq.~(\ref{eq:eh_s_matrixform}). Here $\psing = |0_{2t}\rangle\langle 0_{2t}|$ projects the two-triplet-exciton state onto the singlet sector, where
\begin{equation}
|0_{2t}\rangle = \frac{1}{\sqrt{3}} \left( |1\bar{1}\rangle - |00\rangle + |\bar{1}1\rangle \right)
\end{equation}
is the singlet state of the two triplet excitons.

The rate for the reverse fission process can be obtained analogously or be read off from Eq.\ (\ref{eq:singlet_fission_singlet_basind}),
\begin{equation}
\rate{s \to 2t}{\gs} = - \frac{W'''_{s\to 2t}}{2}\, \Tr \bigg\{ \psing,
  \Big( \one_3 + \tripleta \Big) \otimes \Big( \one_3 + \tripletd \Big) \bigg\} \gs .
\end{equation}

\subsubsection{Point of view of the triplet exciton}

For the diagonal elements of the distribution function, we obtain the three transition rates
\begin{align}
\rate{2t \to s}{\tp} &= - \frac{2 W_{2t \to s} }{3}\, \ftpb \ftma\bracketr{1 + \gs''} , \\
\rate{2t \to s}{\tz} &= - \frac{2 W_{2t \to s} }{3}\, \ftzb \ftza\bracketr{1 + \gs''} , \\
\rate{2t \to s}{\tm} &= - \frac{2 W_{2t \to s} }{3}\, \ftmb \ftpa\bracketr{1 + \gs''} ,
\end{align}
where the factor $2$ is due to the indistinguishable triplet excitons in the initial state. The transition rate referring to the trip\-let-ex\-ci\-ton number density is
\begin{equation}
\rate{2t \to s}{\mto}
=  -\frac{2 \sqrt{6}\, W_{2t \to s}}{9}\,
  \big( \mtob \mtoa - \ctb \cdot \cta + \qtb \cdot \qta \big) (1+\gs'') .
\end{equation}
Since the actual density is $\sqrt{6}\,n_t$, where $n_t$ is the expansion coefficient in Eq.\ (\ref{eq:f_t Q basis}), the transition rate for the actual density is twice the negative transition rate for the sing\-let-ex\-ci\-ton density in Eq.\ (\ref{eq:singlet_fission_singlet}). For the spin transition rate we obtain
\begin{align}
\ratec{2t \to s}{\ct} &= - \frac{W_{2t \to s}}{3}\, \bigg( {-} \frac{2 \sqrt{6}}{3}\, \mtob \cta
  + \frac{2 \sqrt{6}}{3}\, \mtoa \ctb \nonumber \\
&\quad{} - \matqb \cta + \matqa \ctb \bigg) \bracketr{1+\gs''}
\end{align}
and the traceless and symmetric quadrupole transition rate is
\begin{align}
\ratem{2t \to s}{\matq} &=- \frac{W_{2t \to s}}{3}\, \bigg[
  {-} \big( \matq \matqa + \matqa \matq \big)
  + \frac{2}{3} \Tr \big( \matq \matqa \big)\, \one_3 \nonumber \\
&\quad{}- \big( \ct \otimes \cta + \cta \otimes \ct \big)
  + \frac{2}{3}\, \big( \ct \cdot \cta \big)\, \one_3 \nonumber \\
&\quad{}+ \frac{2\sqrt{6}}{3}\, \big( \matqb \mtoa + \mtob \matqa \big)
  \bigg] \bracketr{1 + \gs''} .
\end{align}
The basis-independent transition rate then assumes the form
\begin{equation}
\rate{2t \to s}{\triplet} = - W_{2t \to s} \Tr_{t'} \Big\{ \psing, \tripletb \otimes \tripleta \Big\}
  \bracketr{1 + \gs''},
\end{equation}
where $\Tr_{t'}$ is the partial trace over the sector of the triplet exciton with momentum $\vb{k}'$. The reverse fission process is then described by
\begin{equation}
\rate{s \to 2t}{\triplet} = W_{s\to 2t} \Tr_{t'} \bigg\{ \psing,
  \Big( \one_3 + \tripletb \Big) \otimes \Big( \one_3 + \tripleta \Big) \bigg\} \gs'' .
\end{equation}

\subsection{Absorption of a triplet exciton by an electron}
\label{sub.3G}

The process $e+t\leftrightarrow e$ is more complicated than the previous cases because both particle species carry spin, in contrast to, e.g., $t+s\leftrightarrow t$. For the process $e+t\leftrightarrow e$ to be possible, the total spin of the electron and triplet exciton on the left-hand side must be $1/2$, i.e., they must form a doublet. The two states are
\begin{align}
|\psi^d_\uparrow\rangle &= \frac{\big|\frac{1}{2}\big\rangle |0\rangle
  - \sqrt{2}\, \big|{-}\frac{1}{2}\big\rangle |1\rangle}{\sqrt{3}} , \\
|\psi^d_\downarrow\rangle &= \frac{\sqrt{2}\, \big|\frac{1}{2}\big\rangle |{-}1\rangle
  - \big|{-}\frac{1}{2}\big\rangle |0\rangle}{\sqrt{3}} .
\end{align}
These relations allow us to read off the numerical prefactors in the rates.

\subsubsection{Point of view of the triplet excitons}
\label{subsub.et_to_e.viewt}

In the limit of $z$ polarization, the three transition rates describing triplet excitons are
\begin{align}
\rate{e+t \to e}{\tp} &= - \frac{2 W_{e+ t \to e}}{3}\, \fedowna \ftp (1-\feupc) , \\
\rate{e+t \to e}{\tz} &= - \frac{W_{e+ t \to e}}{3}\, \big[ \feupa  \ftz  (1 - \feupc)
   + \fedowna \ftz  (1 - \fedownc) \big] , \\
\rate{e+t \to e}{\tm} &= - \frac{2 W_{e+ t \to e}}{3}\, \feupa  \ftm  (1 - \fedownc) .
\end{align}
The $\mathrm{SU}(2)$-invariant transition rates in terms of multipole coefficients are obtained in analogy to the previous cases. For the triplet-exciton number density we find
\begin{align}
\rate{e+t \to e}{\mto} &= -\frac{W_{e+ t \to e}}{18}\, \big( 6 \mto \mea
  - 2 \sqrt{6}\, \ct \cdot \cea \nonumber \\
&\quad{} - 3 \mea \mec \mto
  + \mto \cea \cdot \cec \nonumber \\
&\quad{} + \sqrt{6}\, \ct \cdot \cea \mec
  - \sqrt{6}\, \ct \cdot \cec \mea
  + \sqrt{6}\, \cea \cdot \matq \cec \big) ,
\end{align}
for the spin density
\begin{align}
\ratec{e+t \to e}{\ct} &= -\frac{W_{e+ t \to e}}{18}\, \bigg[ 6 \ct \mea
  - 2 \sqrt{6}\, \mto \cea
  - 3 \matq \cea \nonumber \\
&\quad{} + \sqrt{6}\, \mto\, ( \cea \mec - \mea \cec ) \nonumber \\
&\quad{} + \frac{3}{2}\, ( \ct \cdot \cea ) \cec
  + \frac{3}{2}\, ( \ct \cdot \cec ) \cea \nonumber \\
&\quad{} - 3 \ct \mea \mec
  + \frac{3}{2}\, \mec \matq \cea - \frac{3}{2}\, \mea \matq \cec \bigg] ,
\end{align}
and for the symmetric traceless quadrupole density
\begin{align}
&\ratem{e+t \to e}{\matq} = - \frac{W_{e+ t \to e}}{18} \notag \\
&{}\times \bigg( {-} 3\, \bigg[ \cea \otimes \ct + \ct \otimes \cea
  - \frac{2}{3}\, (\cea \cdot \ct)\, \one_3 \bigg] + 6 \mea \matq \nonumber \\
&\quad{} - 3 \mea \mec \matq \nonumber \\
&\quad{}+\sqrt{6} \bigg[ \cea \otimes \cec + \cec \otimes \cea
  - \frac{2}{3}\, ( \cea \cdot \cec)\, \one_3 \bigg] \mto \nonumber \\
&\quad{}+ \frac{3}{2}\, \bigg[ \cea \otimes \ct + \ct \otimes \cea
  - \frac{2}{3}\, ( \cea \cdot \ct )\, \one_3 \bigg] \mec \nonumber \\
&\quad{}- \frac{3}{2}\, \bigg[ \ct \otimes \cec + \cec \otimes \ct
  - \frac{2}{3}\, ( \ct \cdot \cec )\, \one_3 \bigg] \mea \nonumber \\
&\quad{}+ 3 \bigg[ ( \cea \cdot \cec ) \matq
  + \frac{1}{2}\, \cea \otimes (\matq \cec) + \frac{1}{2}\, (\matq \cec) \otimes \cea
  \nonumber \\
&\qquad{} - \frac{1}{2}\, \cec \otimes (\matq \cea) - \frac{1}{2}\, (\matq \cea) \otimes \cec
  \nonumber \\
&\qquad{}- (\cea \otimes \cec) \matq - \matq (\cec \otimes \cea)
  + \frac{2}{3}\, \cea \cdot \matq \cec \, \one_3 \bigg] \bigg) .
\end{align}
Like for the process $e+h\to t$, one might guess at this $\mathrm{SU}(2)$-invariant form based on the limit of \textit{z}-polarized spins but cannot infer it rigorously. The problem lies in the terms containing three $\mathbf{c}$ vectors and in the terms with two $\mathbf{c}$ vectors and the triplet-exciton quadrupole tensor $\matq$. The above expressions have thus been checked by comparing them with the basis-independent expression describing this process, which takes the form
\begin{align}
\rate{e+t \to e}{\triplet} &= -\frac{W_{e+t \to e}}{2}\,
  \Tr_e \bigg[ \Big( \one_2 - \electronc \Big)_6 \pdoub \Big( \electrona \otimes \triplet \Big)
  \nonumber \\
&\quad{} + \Big( \electrona \otimes \triplet \Big) \pdoub \Big( \one_2 - \electronc \Big)_6 \bigg] ,
\end{align}
where $\pdoub$ is the projection operator onto the doublet subspace. This is a $6 \times 6$ matrix. The notation $\bracketr{\bullet}_6$ means that a $2\times 2$ matrix on the doublet subspace is extended to the full two-particle space by adding a $4\times 4$ null matrix on the quartet subspace. The matrix is then transformed into the product basis for electron and triplet-exciton spin to be consistent with $\electron \otimes \triplet$.

\subsubsection{Point of view of the electron}

The perspective of the electron leads to in-scattering and out-scattering terms. The resulting transition rates for the two spin orientations are
\begin{align}
&\rate{e+t \to e}{e \uparrow} = -\frac{W_{e+t \to e} }{3} \nonumber \\
&\quad{} \times \big[ 2\feup  \ftma  (1 - \fedownc) + \feup  \ftza  (1- \feupc) \big] \nonumber \\
&{}+ \frac{W'''_{e+t \to e}}{3}\, \big[ \feupa  \ftzd  (1- \feup) + 2 \fedowna  \ftpd  (1 - \feup) \big] , \\
&\rate{e+t \to e}{e \downarrow} = -\frac{W_{e+t \to e}}{3} \nonumber \\
&\quad{} \times \big[ 2\fedown  \ftpa  (1 - \feupc) + \fedown  \ftza  (1- \fedownc) \big] \nonumber \\
&{}+ \frac{W'''_{e+t \to e}}{3}\, \big[ \fedowna  \ftzd  (1- \fedown) + 2\feupa  \ftmd  (1 - \fedown) \big] ,
\end{align}
where the Clebsch-Gordan coefficients have been used. As the in-scattering and out-scattering terms have the same structure it is sufficient to investigate out-scat\-ter\-ing. For the multipole transition rates this yields
\begin{align}
\rate{e+t \to e}{\me,\,\mathrm{out}} &= -\frac{W_{e+t \to e}}{18}\, \big(
  6 \sqrt{6}\, \me \mtoa
  - 12 \ce \cdot \cta \nonumber \\
&\quad{} - 3 \sqrt{6}\, \me \mtoa \mec
  - 6 \me \cec \cdot \cta
  + \sqrt{6}\, \mtoa \ce \cdot \cec \nonumber \\
&\quad{} + 6 \mec \ce \cdot \cta
  + 6 \ce \cdot \matqa \cec \big) , \\
\ratec{e+t \to e}{\ct,\,\mathrm{out}} &= - \frac{W_{e+t \to e}}{18}\, \bigg[
  12 \me \cta
  - 6 \sqrt{6}\, \ce \mtoa \nonumber \\
&\quad{} - \sqrt{6}\, \me \mtoa \cec
  - 6 \me \cta \mec
  + 3 \sqrt{6}\, \ce \mtoa \mec \nonumber \\
&\quad{} + 6 \ce ( \cta \cdot \cec )
  - \frac{3}{2}\, \me \matqa \cec \bigg] ,
\end{align}
which can be combined to obtain the full transition rate from the point of view of the electron.

Together with the in-scattering terms the basis-in\-de\-pen\-dent form is then
\begin{align}
\rate{e+t \to e}{\electron} &=
  \frac{W'''_{e+t \to e}}{2}\, \Tr_{t'''} \bigg[ \Big(\one_2 - \electron\Big)_6 \pdoub
  \Big(\electrona \otimes \tripletd\Big) \nonumber \\
&\quad{} + \Big(\electrona \otimes \tripletd\Big) \pdoub \Big(\one_2 - \electron\Big)_6 \bigg] \notag \\
&\quad{} - \frac{W_{e+t \to e}}{2}\, \Tr_{t'} \bigg[ \Big(\one_2 - \electronc\Big)_6 \pdoub
  \Big(\electron \otimes \tripleta\Big) \nonumber \\
&\quad{} + \Big(\electron \otimes \tripleta\Big) \pdoub \Big(\one_2 - \electronc\Big)_6 \bigg] .
\end{align}
The absorption and emission of a triplet exciton by a hole is analogous.

\subsection{Fusion of triplet excitons into a triplet exciton}
\label{sub.3H}

The fusion of two triplet excitons into a triplet exciton \cite{StU89} is the most complex process considered here. In the case of $z$ polarization, the transition terms of the components of the density matrix read as
\begin{align}
\label{eq:2t_to_t_zpolarized_1}
&\rate{2t \to t}{\tp} = W'''_{2t \to t}\, \frac{\ftpd \ftza}{2}\, (1 + \ftpb) \notag \\
&\quad{} -2 W_{2t \to t} \bigg[ \frac{\ftpb \ftma}{2}\, (1 + \ftzc)
  + \frac{\ftpb \ftza }{2}\, (1 + \ftpc) \bigg] , \\
&\rate{2t \to t}{\tz} = W'''_{2t \to t}\, \frac{\ftpa \ftmd }{2}\, (1 + \ftz) \notag \\
&\quad{} - 2W_{2t \to t} \bigg[ \frac{\ftpa \ftzb }{2}\, (1 + \ftpc)
  + \frac{\ftma \ftzb }{2}\, (1 + \ftmc) \bigg] , \\
&\rate{2t \to t}{\tm} = W'''_{2t \to t}\, \frac{\ftmd \ftza}{2}\, (1 + \ftmb) \notag \\
&\quad{} - 2W_{2t \to t} \bigg[ \frac{\ftpa \ftmb}{2}\, (1 + \ftzc)
  + \frac{\ftmb \ftza}{2}\, (1 + \ftmc) \bigg] .
\label{eq:2t_to_t_zpolarized_2}
\end{align}
In this process, the in-scattering and out-scattering terms differ in structure. This makes it necessary to investigate both terms separately.

\subsubsection{In-scattering for $2t \to t$}

Combining Eqs.\ (\ref{eq:2t_to_t_zpolarized_1})--(\ref{eq:2t_to_t_zpolarized_2}), we can obtain the transition rates expressed in the multipole expansion. Their generalization to the $\mathrm{SU}(2)$-invariant form is ambiguous at first glance. However, the $\mathrm{SU}(2)$-invariant form can be obtained in analogy to the formation of triplet excitons out of free charge carriers. We construct the density matrix $\tripleta \otimes \tripletd$ on the product space of the two triplet-exciton spins and project it onto the triplet subspace. The resulting $3 \times 3$ matrix, which will be referred to as $\widehat{f}_{tt}$, can be expressed in terms of coefficients of the nine basis matrices $\widehat{S}_i$, $i=1,2,3$, and $\widehat{Q}_j$, $j=1,2,3,4,5$.
For a spin along the $z$ axis, only three coefficients do not vanish:
\begin{align}
n_{tt} &= \frac{2 \sqrt{6}\, \mtoa \mtod-\sqrt{6}\, \ctza \ctzd-\sqrt{6}\, \qtzra \qtzrd}{6} , \\
c_{tt}^z &= \frac{\sqrt{6}\, \ctza \mtod-2 \sqrt{3}\, \ctza \qtzrd+\sqrt{6}\, \ctzd \mtoa
  -2 \sqrt{3}\, \ctzd \qtzra}{6} , \\
q_{tt5} &= \frac{2 \sqrt{2}\, \ctza \ctzd-2 \mtoa \qtzrd-2 \mtod \qtzra-2 \sqrt{2}\, \qtzra \qtzrd}{6} .
\end{align}
Combining these equations and using Eqs.\ (\ref{eq:2t_to_t_zpolarized_1})--(\ref{eq:2t_to_t_zpolarized_2}), we can unambiguously generalize the rates to their $\mathrm{SU}(2)$-invariant form:
\begin{align}
\rate{2t \to t}{\mto,\,\mathrm{in}} 
  &= W'''_{2t \to t} \bigg[ \mtott + \frac{\sqrt{6}}{3}\, \mtob \mtott
  + \frac{\sqrt{6}}{3}\, \ct \cdot \ctt \nonumber \\
&\quad{}+ \frac{\sqrt{6}}{3}\, \qtb \cdot \qtt \bigg] , \\
\ratec{2t \to t}{\ct,\,\mathrm{in}} &= W'''_{2t \to t} \bigg[ \ctt + \frac{\sqrt{6}}{3}\, \mtott \cta
  + \frac{\sqrt{6}}{3}\, \ctt \mtoa \nonumber \\
&\quad{}+ \frac{1}{2}\, \matqtt \cta + \frac{1}{2}\, \matqa \ctt \bigg] , \\
\ratem{2t \to t}{\matq,\,\mathrm{in}} &= W'''_{2t \to t}
  \bigg[ \matqtt + \frac{\sqrt{6}}{3}\, \mtott \matqb + \frac{\sqrt{6}}{3}\, \mtob \matqtt \nonumber \\
&\quad{}- \frac{1}{2}\, \matqtt \matqb - \frac{1}{2}\, \matqb \matqtt
  + \frac{1}{3}\, \Tr \big(\matqtt \matqb\big) \one_3 \nonumber \\
&\quad{}- \frac{1}{3}\, (\ctt \cdot \ctb) \one_3
  + \frac{1}{2}\, (\ctt \otimes \ctb + \ctb \otimes \ctt) \bigg] .
\end{align}
The basis-independent in-scattering term reads as
\begin{align}
\rate{2t \to t}{\triplet,\,\mathrm{in}} &= \frac{W'''_{2t \to t}}{2}\,
  \Tr_{t'''} \bigg[ \Big(\one_3 + \tripletb\Big)_9 \ptrip \Big(\tripleta \otimes \tripletd\Big) \nonumber \\
&\quad{} + \Big(\tripleta \otimes \tripletd\Big) \ptrip \Big(\one_3 + \tripletb\Big)_9 \bigg] .
\end{align}
The notation $(\bullet)_9$ is analogous to $(\bullet)_4$ in Sec.\ \ref{subsub.eh_to_t.viewe} and to $(\bullet)_6$ in Sec.~\ref{subsub.et_to_e.viewt}.

\subsubsection{Out-scattering for $2t \to t$}

For out-scattering, the strategy employed for in-scattering does not work since the rates for the case of $z$ polarization are not symmetric in the two incoming triplet excitons at the outer momentum $\vb{k}$ and the running momentum $\vb{k}'$. Hence, the projected density matrix on the product space, $\widehat{f}_{tt}$, is not useful. The situation is analogous to the process $e+h \to t$ from the point of view of the electron or hole; see Sec.\ \ref{subsub.eh_to_t.viewe}. Like for $e + h \to t$, the matrix-valued transition rate in terms of projection operators and density matrices
\begin{align}
\rate{2t \to t}{\triplet,\,\mathrm{out}} &= -W_{2t \to t}\, \Tr_{t'} \bigg[
  \Big( \one_3 + \tripletc \Big)_9 \ptrip \Big( \tripletb \otimes \tripleta \Big) \nonumber \\
&\quad{} + \Big( \tripletb \otimes \tripleta \Big) \ptrip \Big( \one_3 + \tripletc \Big)_9 \bigg]
\label{eq:2t.t.nobasis}
\end{align}
can be used to read off the correct $\mathrm{SU}(2)$-invariant multipole transition rates. For the triplet exciton number density this yields
\begin{widetext}
\begin{align}
\rate{2t \to t}{\mto,\,\mathrm{out}} &= - \frac{W_{2t \to t}}{3}\, \bigg[ 2 \sqrt{6}\, \mtob \mtoa
  - \sqrt{6}\, \ctb \cdot \cta - \frac{\sqrt{6}}{2}\, \Tr \matqb \matqa \nonumber \\
&\quad{} + 4 \mtob \mtoa \mtoc
  + 2 \big( \mtob \cta \cdot \ctc + \mtoa \ctb \cdot \ctc + \mtoc \ctb \cdot \cta \big)
  - \Big( \mtob \Tr  \matqa \matqc + \mtoa \Tr \matqb \matqc
    + \mtoc \Tr \matqb \matqa \Big) \nonumber \\
&\quad{}- \sqrt{6}\, \Big( \ctb \cdot \matqa \ctc + \cta \cdot \matqb \ctc
    + \ctb \cdot \matqc \cta \Big)
  - \frac{\sqrt{6}}{2}\, \Big(\Tr \matqb \matqa \matqc
    + \Tr \matqb \matqc \matqa \Big) \bigg] .
\end{align}
The spin transition rate becomes
\begin{align}
\ratec{2t \to t}{\ct,\,\mathrm{out}} &= -\frac{W_{2t \to t}}{6}\, \Big[
  2\sqrt{6}\, (- 2 \mtob \cta + \mtoa \ctb)
  - 3 \Big( \matqb \cta + \matqa \ctb \Big)
  - 4 \big( \mtob \mtoc \cta -\mtob \mtoa \ctc -2 \mtob \mtoc \cta \big) \nonumber \\
&\quad{} - \sqrt{6}\, \Big( \mtoa \matqc \ctb
  - \mtoa \matqb \ctc + \mtoc \matqa \ctb
  + \mtoc \matqb \cta + 2 \mtob \matqa \ctc - 2 \mtob \matqc \cta \Big)
  \nonumber \\
&\quad{} - 3 \big( \cta (\ctb \cdot \ctc) + \ctc (\cta \cdot \ctb) - 2 \ctb (\cta \cdot \ctc) \big)
  - 3 \Big( \matqa \matqc \ctb + \matqc \matqa \ctb
  + \matqb \matqa \ctc - \matqb \matqc \ctb \Big) \Big] .
\end{align}
The quadrupole part reads as
\begin{align}
\ratem{2t \to t}{\matq,\,\mathrm{out}} &= - \frac{W_{2t \to t}}{6}\, \bigg( 2 \sqrt{6}\,
  \big( - \mtob \matqa + 2 \mtoa \matqb )
  - 3 \big[ \ctb \otimes \cta + \cta \otimes \ctb - 2 (\cta \cdot \ctb) \one_3 \big]
  - 3 \big( \matqb \matqa + \matqa \matqb \big)
  - 2 \Tr \big( \matqb \matqa \big) \one_3 \nonumber \\
&\quad{} - 4 \Big( \mtob \mtoa \matqc + \mtob \mtoc \matqa - 2 \mtoa \mtoc \matqb \Big) \nonumber \\
&\quad{} - \sqrt{6}\, \big[ 2 \mtob ( \cta \otimes \ctc + \ctc \otimes \cta )
    - \mtoa ( \ctb \otimes \ctc + \ctc \otimes \ctb )
    + \mtoc ( \ctb \otimes \cta + \cta \otimes \ctb ) \big] \nonumber \\
&\quad{} + \frac{2 \sqrt{6}}{3}\, \Big[ 2 \mtob (\cta \cdot \ctc) - \mtoa (\ctb \cdot \ctc)
    + \mtoc (\ctb \cdot \cta) \Big] \one_3 \nonumber \\
&\quad{} - \sqrt{6}\, \Big[ 2 \mtob \Big( \matqa \matqc + \matqc \matqa \Big)
    - \mtoa \Big( \matqb \matqc + \matqc \matqb \Big)
    - \mtoc \Big( \matqb \matqa + \matqa \matqb \Big) \Big] \nonumber \\
&\quad{} + \frac{2 \sqrt{6}}{3}\, \Big[ 2 \mtob \Tr \big( \matqa \matqc \big)
    - \mtoa \Tr \big( \matqb \matqc \big)
    - \mtoc \Tr \big( \matqb \matqa \big) \Big] \one_3 \nonumber \\
&\quad{} - 3 \Big[ \big( \ctb \otimes \ctc \big) \matqa + \matqa \big( \ctc \otimes \ctb \big)
    - \big( \ctb \otimes \cta \big) \matqc - \matqc \big( \cta \otimes \ctb \big)
    - \big( \cta \otimes \ctc \big) \matqb - \matqb \big( \ctc \otimes \cta \big)
      \nonumber \\
&\qquad{} - \big( \ctc \otimes \cta \big) \matqb - \matqb \big( \cta \otimes \ctc \big) \Big]
  - 2 \Big( \ctb \cdot \matqc \cta- \ctb \cdot \matqa \ctc
    + \ctc \cdot \matqb \cta \Big) \one_3 \nonumber \\
&\quad{} + 3 \Big( \matqb \matqa \matqc + \matqc \matqa \matqb
    + \matqa \matqc \matqb + \matqb \matqc \matqa \Big)
  - 4 \Tr \Big( \matqb \matqa \matqc \Big) \one_3
    - 6 \Tr \Big(\matqa \matqc\Big) \matqb \bigg) .
\end{align}
\end{widetext}
The derivation of the rates for the spin and for the quadrupole moment again require the basis-independent form in Eq.\ (\ref{eq:2t.t.nobasis}).

The reverse fission process $t \to 2t$ can be investigated in the same way. In the out-scattering term, the introduction of $\widehat{f}_{1 - tt}$ leads to the same structure as in the in-scattering term for $2t \to t$.

\section{Examples}
\label{sec.example}

In this section, we illustrate the theoretical framework by applying it to a simple model. In order to reduce numerical complications, we consider a one-dimensional toy model containing fission and fusion processes, disorder scattering of singlet and triplet excitons, as well as relaxation of singlet and triplet excitons. All bare rates $W$ are assumed to be independent of momentum, except as dictated by scattering being elastic.
The temperature is assumed to be sufficiently low so that the equilibrium thermal population of excitonic states is negligible. A magnetic field, if present, is taken to be uniform, constant in time, and oriented along the \textit{z} direction.

The dispersion relations of singlet and triplet excitons are written as
\begin{align}
\epsilon_s(k) &= \Delta \epsilon + \frac{\hbar^2k^2}{2m_s} , \\
\epsilon_t(k,m) &= \frac{\hbar^2k^2}{2m_t} - g_\mathrm{tr} \mu_B B m ,
\end{align}
respectively, where $m=1,0,-1$ specifies the spin state of triplet excitons and
\begin{equation}
\Delta \epsilon \equiv \frac{\hbar^2k_0^2}{2m_t}
\end{equation}
is the energy difference between singlet and triplet excitons, which defines a momentum scale $k_0$.

Choosing our units such that $\hbar/m_t=1$ and defining the parameters
\begin{align}
\alpha &\equiv \frac{m_t}{m_s} = \frac{\hbar}{m_s} , \\
\omega_L &\equiv \frac{1}{\hbar}\, g_\mathrm{tr} \mu_B B ,
\end{align}
the Boltzmann equation for the singlet-exciton distribution function $f_s(x,k,t)$ can be written as
\begin{equation}
\frac{\partial f_s}{\partial t} + \alpha k\, \frac{\partial f_s}{\partial x} = I^{f_s} ,
\end{equation}
and the corresponding equation for the matrix-valued triplet-exciton distribution function $\widehat f_t(x,k,t)$ as
\begin{equation}
\frac{\partial \widehat f_t}{\partial t} + k\, \frac{\partial \widehat f_t}{\partial x}
  + i \omega_L\, [\widehat S_3, \widehat f_t] = \widehat I^{\widehat f_t} .
\label{eq.example.LHStriplet}
\end{equation}
Here $I^{f_s}$ and $\widehat I^{\widehat f_t}$ are the full scattering integrals, which contain the following contributions, where we only make the momentum dependence explicit:

(a) Decay is discussed in Sec.\ \ref{sub.3A}. It is described by the scattering integrals
\begin{align}
I^{f_s}_{s\to \ket{0}}(k) &= - \widetilde W_{s\to \ket{0}}\, f_s(k) \equiv - \eta_{s0}\, f_s(k) , \\
\widehat I^{\widehat f_t}_{t\to \ket{0}}(k) &= - \widetilde W_{t\to \ket{0}}\, \widehat f_t(k)
  \equiv - \eta_{t0}\, \widehat f_t(k) .
\end{align}
\phantom{x}

(b) Disorder scattering is discussion in the introduction to Sec.\ \ref{sec.scattering} and is described by
\begin{align}
I^{f_s}_{s\to s} &= \int \frac{dk'}{2\pi}\, w_{s\to s'}\, \delta\bigg( \frac{\hbar^2 (k')^2}{2m_s}
  - \frac{\hbar^2 k^2}{2m_s} \bigg)\, [ f_s(k') - f_s(k) ] \nonumber \\*[0ex]
&= w_{s\to s'}\, \frac{m_s}{2\pi \hbar^2}\, \frac{f_s(-k) - f_s(k)}{|k|} \nonumber \\*[0ex]
&\equiv \eta^\mathrm{dis}_s\, \frac{f_s(-k) - f_s(k)}{|k|} , \\
\widehat I^{\widehat f_t}_{t\to t} &= \int \frac{dk'}{2\pi}\, w_{t\to t'}\,
  \delta\bigg( \frac{\hbar^2 (k')^2}{2m_t} - \frac{\hbar^2 k^2}{2m_t} \bigg)\,
  \Big[ \widehat f_t(k') - \widehat f_t(k) \Big] \nonumber \\
&= w_{t\to t'}\, \frac{m_t}{2\pi \hbar^2}\, \frac{\widehat f_t(-k) - \widehat f_t(k)}{|k|}
  \nonumber \\
&\equiv \eta^\mathrm{dis}_t\, \frac{\widehat f_t(-k) - \widehat f_t(k)}{|k|} .
\end{align}
Note that the singlet-triplet energy difference $\Delta\epsilon$ and the magnetic energy of triplet excitons cancel.

(c) Fusion is discussion in Sec.\ \ref{sub.3F}. From the point of view of the singlet excitons, fusion is described by
\begin{widetext}
\begin{align}
I^{f_s}_{2t\to s}(k) &= \int \frac{dk'}{2\pi}\, \frac{w_{2t\to s}}{2}\,
  \delta\bigg( \frac{\hbar^2(k')^2}{2m_t} + \frac{\hbar^2(k-k')^2}{2m_t} - \Delta\epsilon
    - \frac{\hbar^2k^2}{2m_s} \bigg)
  \Tr \big\{ \psing, \widehat f_t(k') \otimes \widehat f_t(k-k') \big\}\, [ 1 + f_s(k) ] \nonumber \\
&= w_{2t\to s}\, \frac{m_t}{4\pi\hbar^2}\, \frac{1}{\kappa}\, [ 1 + f_s(k) ]
  \bigg( \Tr \psing \bigg[ \widehat f_t\bigg( \frac{k}{2} + \kappa \bigg)
    \otimes \widehat f_t\bigg( \frac{k}{2} - \kappa \bigg) \bigg]
  + \Tr \psing \bigg[ \widehat f_t\bigg( \frac{k}{2} - \kappa \bigg)
    \otimes \widehat f_t\bigg( \frac{k}{2} + \kappa \bigg) \bigg] \bigg) \nonumber \\
&\equiv w_\mathrm{fusion}\, \frac{1}{\kappa}\, [ 1 + f_s(k) ]
  \bigg( \Tr \psing \bigg[ \widehat f_t\bigg( \frac{k}{2} + \kappa \bigg)
    \otimes \widehat f_t\bigg( \frac{k}{2} - \kappa \bigg) \bigg]
  + \Tr \psing \bigg[ \widehat f_t\bigg( \frac{k}{2} - \kappa \bigg)
    \otimes \widehat f_t\bigg( \frac{k}{2} + \kappa \bigg) \bigg] \bigg) ,
\end{align}
with
\begin{equation}
\kappa \equiv \sqrt{\textstyle k_0^2 + \frac{m_t}{2m_s}\, k^2 - \frac{k^2}{4}}
  = \sqrt{\textstyle k_0^2 + \frac{\alpha}{2}\, k^2 - \frac{k^2}{4}} ,
\label{eq.example.kappa}
\end{equation}
if $\kappa$ is real, otherwise the scattering integral vanishes. Note that the magnetic energy of triplet excitons cancels. Introducing the unitary part of the antiunitary time-reversal operator,
\begin{equation}
\widehat U_T = e^{i \widehat S_2 \pi} = \begin{pmatrix}
    0 & 0 & 1 \\ 0 & -1 & 0 \\ 1 & 0 & 0
  \end{pmatrix} = \widehat U_T^\dagger ,
\end{equation}
we can simplify the expression to
\begin{equation}
I^{f_s}_{2t\to s}(k) = \frac{2}{3}\, w_\mathrm{fusion}\, \frac{1}{\kappa}\, [ 1 + f_s(k) ]
  \Tr \widehat f_t\bigg( \frac{k}{2} + \kappa \bigg) \widehat U_T
    \widehat f_t^{\,T}\bigg( \frac{k}{2} - \kappa \bigg) \widehat U_T^\dagger .
\end{equation}
The corresponding scattering integral for the triplet excitons reads as
\begin{align}
\widehat I^{\widehat f_t}_{2t\to s} &= - \int \frac{dk'}{2\pi}\, w_{2t\to s}\,
  \delta\bigg( \frac{\hbar^2k^2}{2m_t} + \frac{\hbar^2(k')^2}{2m_t} - \Delta\epsilon
     - \frac{\hbar^2(k+k')^2}{2m_s} \bigg)
  \Tr_{t'} \Big\{ \psing, \widehat f_t(k) \otimes \widehat f_t(k') \Big\}\,
  [ 1 + f_s(k+k') ] \nonumber \\
&= - \frac{2 w_\mathrm{fusion}}{|1-\alpha|}\, \frac{1}{\kappa'}\,
  \bigg( \Tr_{t'} \bigg\{ \psing, \widehat f_t(k)
    \otimes \widehat f_t\bigg( \frac{\alpha}{1-\alpha}\, k + \kappa' \bigg) \bigg\}
    \bigg[ 1 + f_s\bigg( \frac{1}{1-\alpha}\, k + \kappa' \bigg) \bigg] \nonumber \\
&\quad{} + \Tr_{t'} \bigg\{ \psing, \widehat f_t(k)
    \otimes \widehat f_t\bigg( \frac{\alpha}{1-\alpha}\, k - \kappa' \bigg) \bigg\}
    \bigg[ 1 + f_s\bigg( \frac{1}{1-\alpha}\, k - \kappa' \bigg) \bigg] \bigg) ,
\end{align}
with
\begin{equation}
\kappa'
  \equiv \sqrt{\textstyle \frac{2}{1-\alpha}\, \Big( k_0^2 + \frac{\alpha - 1/2}{1-\alpha}\, k^2 \Big)} ,
\label{eq.example.kappap}
\end{equation}
if $\kappa'$ is real, otherwise the scattering integral vanishes. With some algebra, the expression can be rewritten as
\begin{align}
\widehat I^{\widehat f_t}_{2t\to s} &= - \frac{2}{3}\, \frac{w_\mathrm{fusion}}{|1-\alpha|}\,
  \frac{1}{\kappa'}\,
  \bigg( \bigg[ \widehat f_t(k) \widehat U_T \widehat f_t^{\,T}\bigg( \frac{\alpha}{1-\alpha}\, k + \kappa' \bigg)
      \widehat U_T^\dagger
    + \widehat U_T \widehat f_t^{\,T}\bigg( \frac{\alpha}{1-\alpha}\, k + \kappa' \bigg) \widehat U_T^\dagger
      \widehat f_t(k) \bigg] \bigg[ 1 + f_s\bigg( \frac{1}{1-\alpha}\, k + \kappa' \bigg) \bigg] \nonumber \\
&\quad{} + \bigg[ \widehat f_t(k) \widehat U_T
      \widehat f_t^{\,T}\bigg( \frac{\alpha}{1-\alpha}\, k - \kappa' \bigg) \widehat U_T^\dagger
    + \widehat U_T \widehat f_t^{\,T}\bigg( \frac{\alpha}{1-\alpha}\, k - \kappa' \bigg) \widehat U_T^\dagger
      \widehat f_t(k) \bigg] \bigg[ 1 + f_s\bigg( \frac{1}{1-\alpha}\, k - \kappa' \bigg) \bigg] \bigg) ,
\end{align}
which only contains $3\times 3$ instead of $9\times 9$ matrices.

(d) Fission is also discussed in Sec.\ \ref{sub.3F}. From the point of view of the singlet excitons, we have
\begin{align}
I^{f_s}_{s\to 2t} &= - \int \frac{dk'}{2\pi}\, \frac{w_{s\to 2t}}{2}\,
  \delta\bigg( \frac{\hbar^2(k')^2}{2m_t} + \frac{\hbar^2(k-k')^2}{2m_t} - \Delta\epsilon
    - \frac{\hbar^2k^2}{2m_s} \bigg)
  \Tr \Big\{ \psing, \big[ \one_3 + \widehat f_t(k') \big]
    \otimes \big[ \one_3 + \widehat f_t(k-k') \big] \Big\}\, f_s(k) \nonumber \\
&= - w_{s\to 2t}\, \frac{m_t}{4\pi\hbar^2}\, \frac{1}{\kappa}\, f_s(k)\,
  \bigg[ \Tr \psing \bigg( \bigg[ \one_3 + \widehat f_t\bigg( \frac{k}{2} + \kappa \bigg) \bigg]
    \otimes \bigg[ \one_3 + \widehat f_t\bigg( \frac{k}{2} - \kappa \bigg) \bigg] \bigg) \nonumber \\
&\quad{} + \Tr \psing \bigg( \bigg[ \one_3 + \widehat f_t\bigg( \frac{k}{2} - \kappa \bigg) \bigg]
    \otimes \bigg[ \one_3 + \widehat f_t\bigg( \frac{k}{2} + \kappa \bigg) \bigg] \bigg) \bigg] \nonumber \\
&\equiv - w_\mathrm{fission} \frac{1}{\kappa}\, f_s(k)\,
  \bigg[ \Tr \psing \bigg( \bigg[ \one_3 + \widehat f_t\bigg( \frac{k}{2} + \kappa \bigg)
    \otimes \bigg[ \one_3 + \widehat f_t\bigg( \frac{k}{2} - \kappa \bigg) \bigg] \bigg) \bigg] \nonumber \\
&\quad{} + \Tr \psing \bigg( \bigg[ \one_3 + \widehat f_t\bigg( \frac{k}{2} - \kappa \bigg) \bigg]
    \otimes \bigg[ \one_3 + \widehat f_t\bigg( \frac{k}{2} + \kappa \bigg) \bigg] \bigg) \bigg] ,
\end{align}
with $\kappa$ given by Eq.\ (\ref{eq.example.kappa}), if $\kappa$ is real and zero otherwise. The expression can be simplified to
\begin{equation}
I^{f_s}_{s\to 2t} = -\frac{2}{3}\, w_\mathrm{fission}\, \frac{1}{\kappa}\, f_s(k)\,
  \Tr \bigg[ \one_3 + \widehat f_t\bigg( \frac{k}{2} + \kappa \bigg) \bigg] \widehat U_T
    \bigg[ \one_3 + \widehat f_t^{\,T}\bigg( \frac{k}{2} - \kappa \bigg) \bigg] \widehat U_T^\dagger .
\end{equation}
Analogously, we find
\begin{align}
\widehat I^{\widehat f_t}_{s\to 2t} &= \frac{2}{3}\, \frac{w_\mathrm{fission}}{|1-\alpha|}\,
  \frac{1}{\kappa'}\,
  \bigg[ \bigg( \big[ \one_3 + \widehat f_t(k) \big] \widehat U_T
      \bigg[ \one_3 + \widehat f_t^{\,T}\bigg( \frac{\alpha}{1-\alpha}\, k + \kappa' \bigg) \bigg]
      \widehat U_T^\dagger
    + \hat U_T \bigg[ \one_3 + \widehat f_t^{\,T}\bigg( \frac{\alpha}{1-\alpha}\, k + \kappa' \bigg) \bigg]
      \widehat U_T^\dagger \big[ \one_3 + \widehat f_t(k) \big] \bigg) \nonumber \\
&\qquad{} \times  f_s\bigg( \frac{1}{1-\alpha}\, k + \kappa' \bigg) \nonumber \\
&\quad{} + \bigg( \big[ \one_3 + \widehat f_t(k) \big] \widehat U_T
      \bigg[ \one_3 + \widehat f_t^{\,T}\bigg( \frac{\alpha}{1-\alpha}\, k - \kappa' \bigg) \bigg]
      \widehat U_T^\dagger
    + \hat U_T \bigg[ \one_3 + \widehat f_t^{\,T}\bigg( \frac{\alpha}{1-\alpha}\, k - \kappa' \bigg) \bigg]
      \widehat U_T^\dagger \big[ \one_3 + \widehat f_t(k) \big] \bigg) \nonumber \\
&\qquad{} \times f_s\bigg( \frac{1}{1-\alpha}\, k - \kappa' \bigg) \bigg] ,  
\end{align}
\end{widetext}
with $\kappa'$ given by Eq.\ (\ref{eq.example.kappap}), if $\kappa'$ is real and zero otherwise.

One can now expand $\widehat f_t$ into multipoles using Eq.\ (\ref{eq:f_t Q basis}). This is particularly useful if some of the multipoles vanish by symmetry. In any case, the equations to be solved are coupled first-order nonlinear partial differential equations for $f_s$ and $\widehat f_t$ as functions of time $t$ and position $x$. The momentum $k$ acts as a parameter labeling different coupled functions. None of the coefficients depend on $x$ or $t$. Nevertheless, due to the nonlinearity of the equations, an analytical solution seems out of reach. We therefore use numerical forward propagation of the initial functions with discrete time steps and also discretize the $x$ and $k$ dependence.

\begin{figure}
\includegraphics[width=0.75\columnwidth]{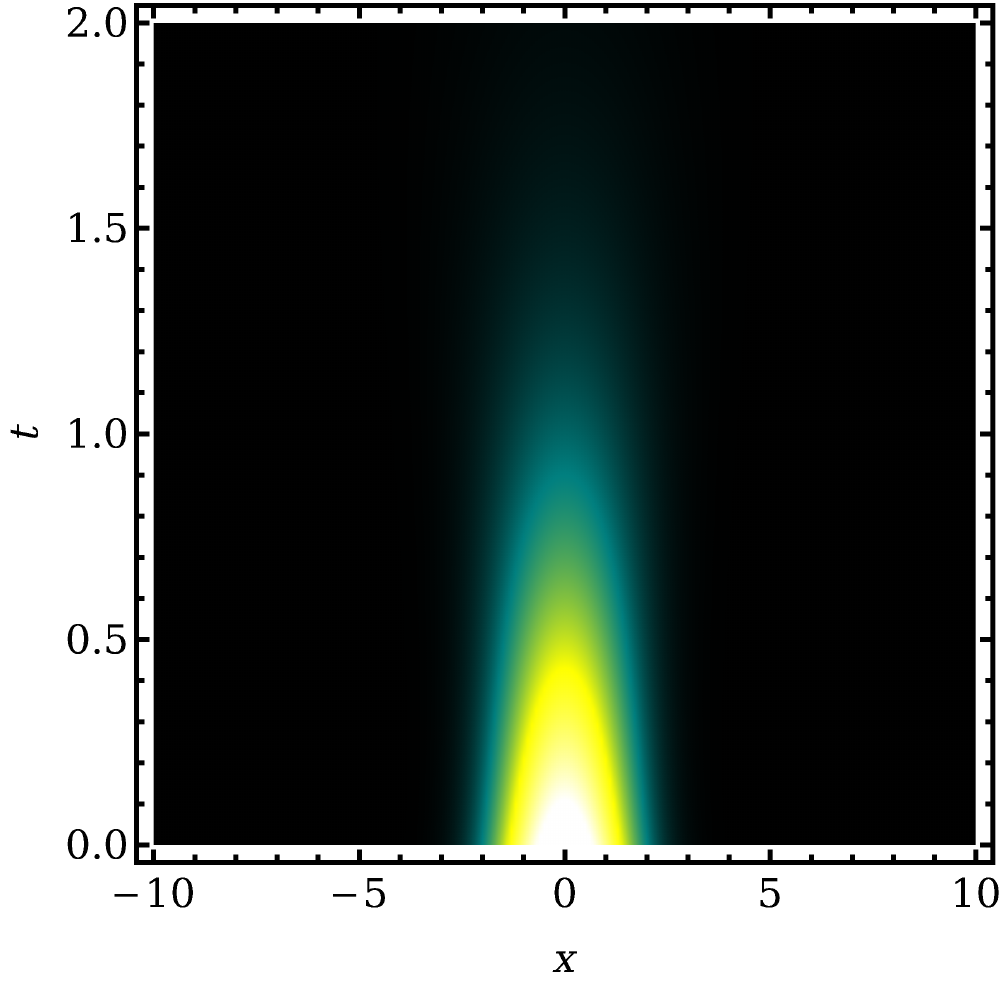}
\caption{Spatial distribution function $\rho_s(x,t)$ for singlet excitons for the case without fission. There are no triplet excitons in this case. The parameters are $\alpha=0.8$, $k_0=4$, $\sigma_k=1$, $\sigma_x=1$, $\eta_{s0}=2$, and all other rates are set to zero. The discretization intervals for momentum, space, and time are $\Delta k=0.1$, $\Delta x=0.1$, and $\Delta t= 2.5\times 10^{-3}$.
\label{fig.NM.1}}
\end{figure}

\begin{figure}
\raisebox{45ex}[0ex][0ex]{(a)}\includegraphics[width=0.75\columnwidth]{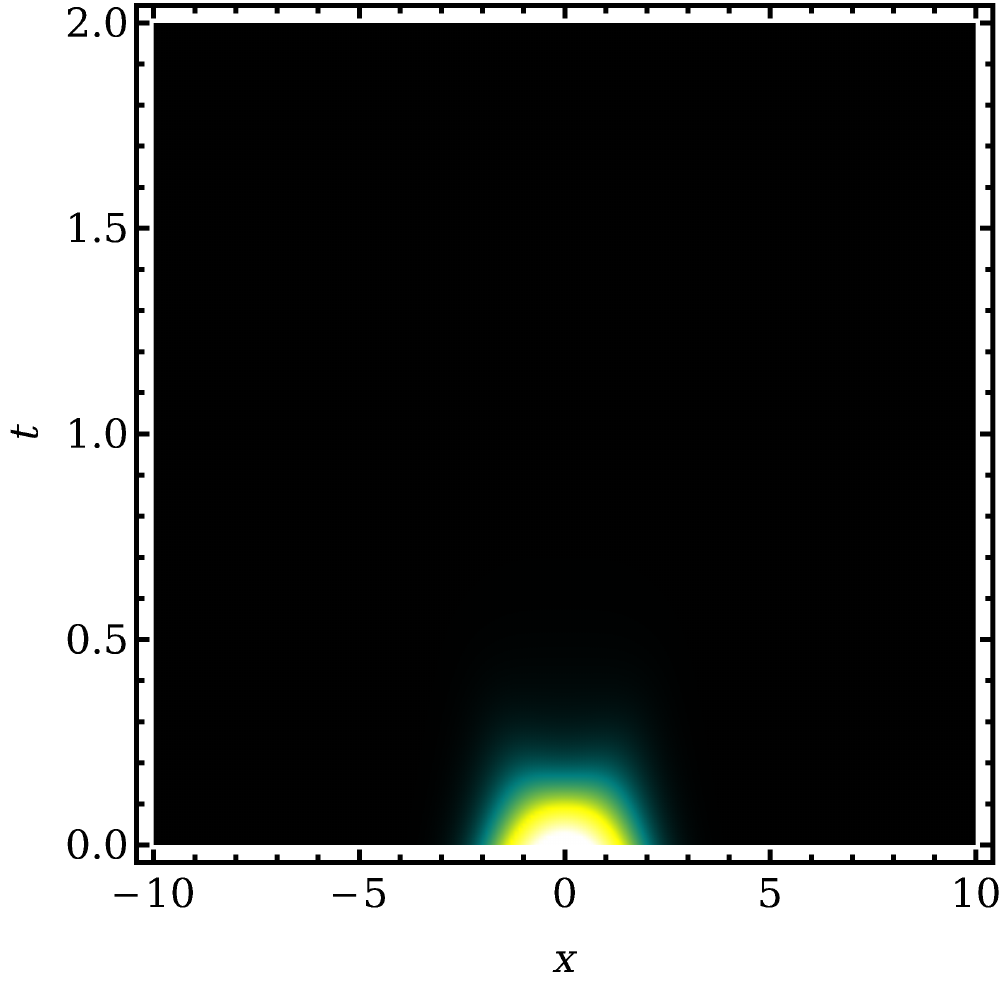}\\[1ex]
\raisebox{45ex}[0ex][0ex]{(b)}\includegraphics[width=0.75\columnwidth]{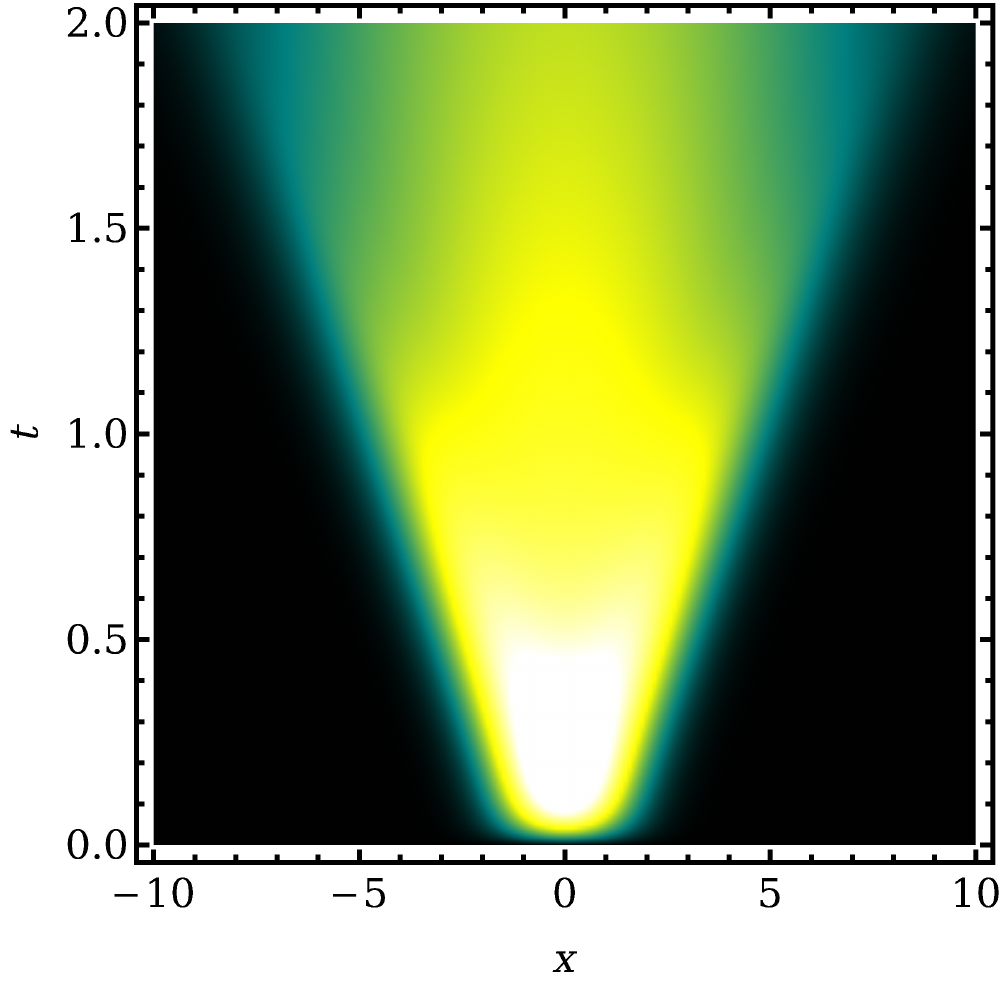}
\caption{Spatial distribution function for (a) singlet and (b) triplet excitons as a function of time for the case with fission but without fusion. The parameters are $\eta_t^\mathrm{dis} = 6$ and $w_{s\to 2t}=10$, the other parameters are the same as in Fig.~\ref{fig.NM.1}.
\label{fig.NM.3}}
\end{figure}

\begin{figure}
\raisebox{45ex}[0ex][0ex]{(a)}\includegraphics[width=0.75\columnwidth]{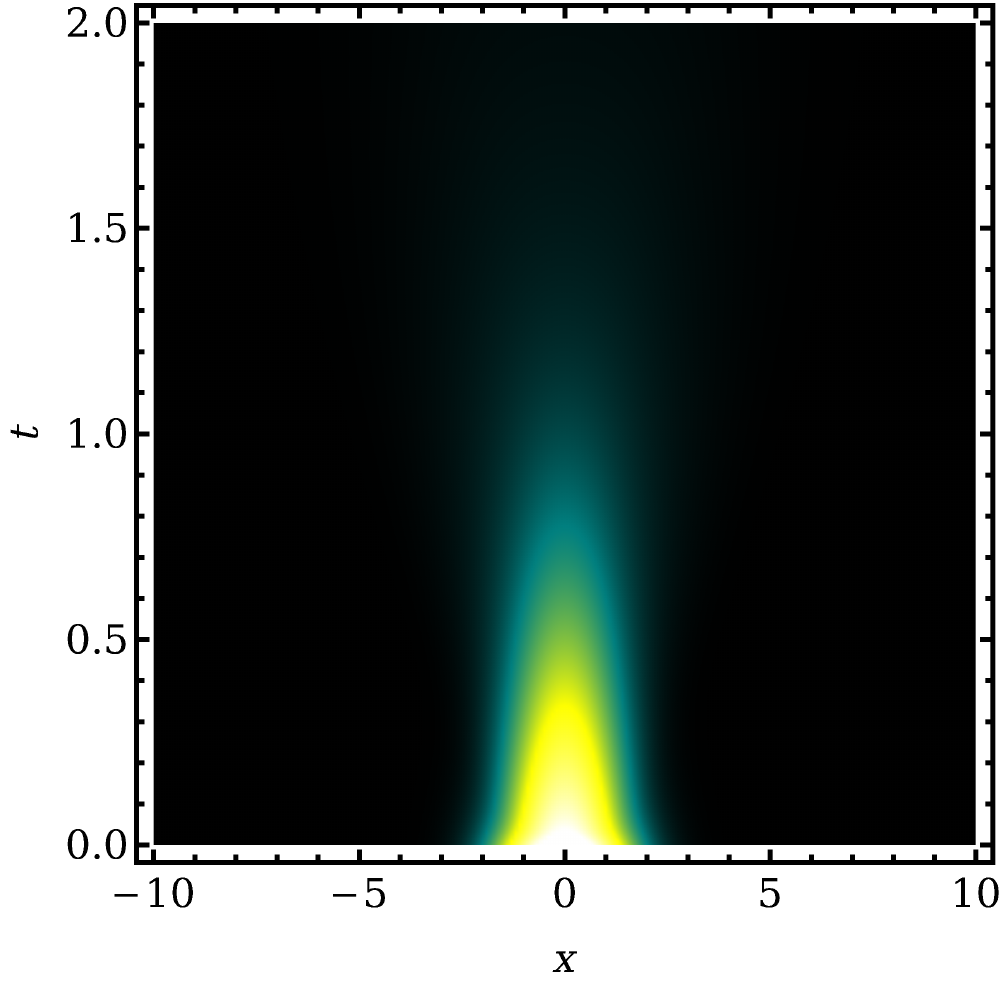}\\[1ex]
\raisebox{45ex}[0ex][0ex]{(b)}\includegraphics[width=0.75\columnwidth]{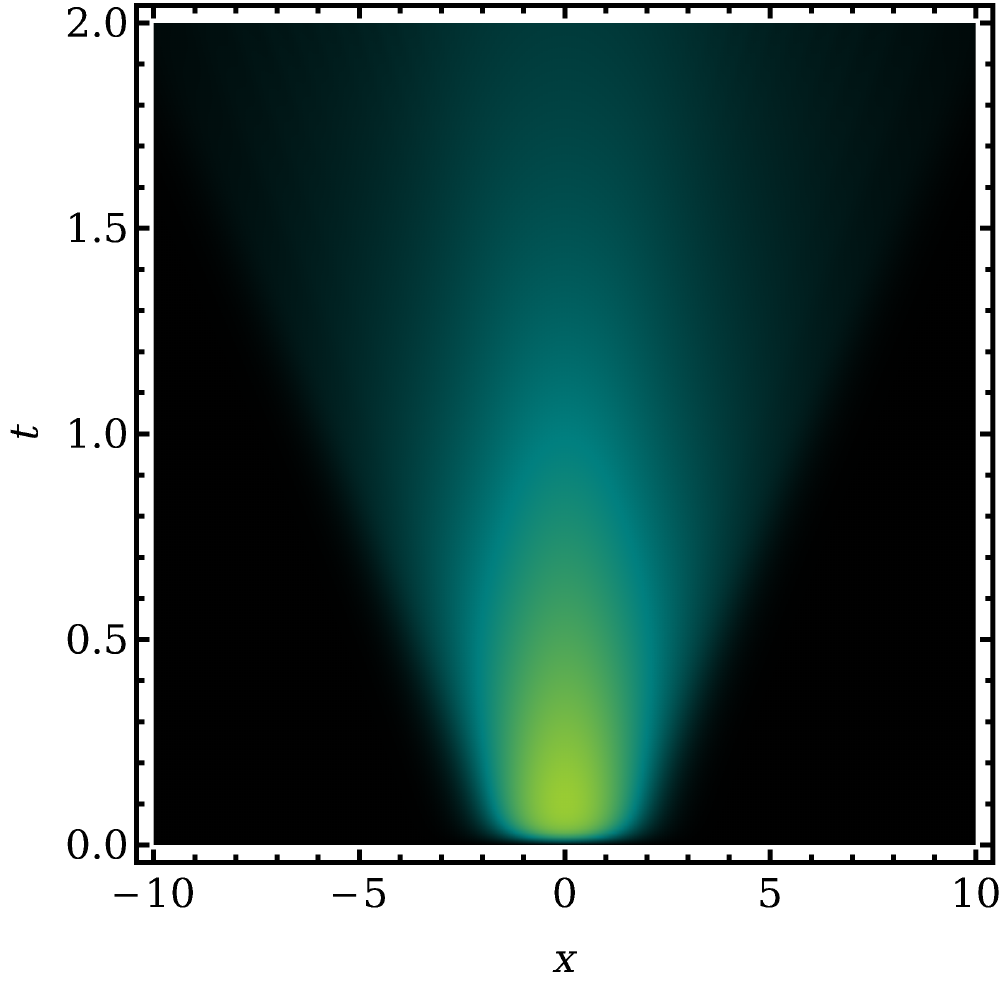}
\caption{Spatial distribution function for (a) singlet and (b) triplet excitons as a function of time for the case with fission and strong fusion. The fusion rate is $w_{2t\to s}=400$, the other parameters are the same as in Fig.~\ref{fig.NM.3}.
\label{fig.NM.4}}
\end{figure}

As an example, we consider a localized distribution of singlet excitons, photogenerated close to the energy threshold. The distribution is modeled by a Gaussian function with mean zero and width $\sigma_x$ in real space multiplied by a Gaussian function with mean zero and width $\sigma_k$ in momentum space. To reflect some of the features of real systems, the singlet excitons are assumed to decay rapidly.
We consider the singlet-excition distribution function in real space,
\begin{equation}
\rho_s(x,t) = \int \frac{dk}{2\pi}\, f_s(k,x,t) ,
\end{equation}
and the distribution function of the occupation of triplet excitons in real space,
\begin{equation}
\rho_t(x,t) = \int \frac{dk}{2\pi}\, \Tr \widehat f_t(k,x,t) .
\end{equation}
For reference, we plot the time evolution of the singlet-exciton distribution function $\rho_s$ in Fig.\ \ref{fig.NM.1} for the situation without fission. In this case, the singlets decay rapidly and triplet excitons are not generated. Note that the densities $\rho_s$ and $\rho_t$ in all plots are presented in arbitrary units but on the same color scale.

If fission (but not fusion) is switched on the singlet excitons quickly transition into triplet excitatons, as shown in Fig.\ \ref{fig.NM.3}(a).
The triplet excitons show predominant diffusive motion due to disorder scattering, see Fig.\ \ref{fig.NM.3}(b). The cone from ballistically moving, i.e., not scattered, excitons is visible as a weaker feature.

If we now also switch on fusion with a high bare rate we observe recovery of singlet excitons and loss of triplet excitons, as expected, see Fig.\ \ref{fig.NM.4}. However, the distribution of recreated singlet excitons is much narrower than the one of the triplet excitons. We attribute this to the fact that for our parameters two triplet excitons must collide head on in order to fuse, which is more likely in the center of the triplet cloud. It is an interesting question for the future to what extent such an effect is also present in real three-dimensional systems.

\begin{figure}
\raisebox{45ex}[0ex][0ex]{(a)}\includegraphics[width=0.75\columnwidth]{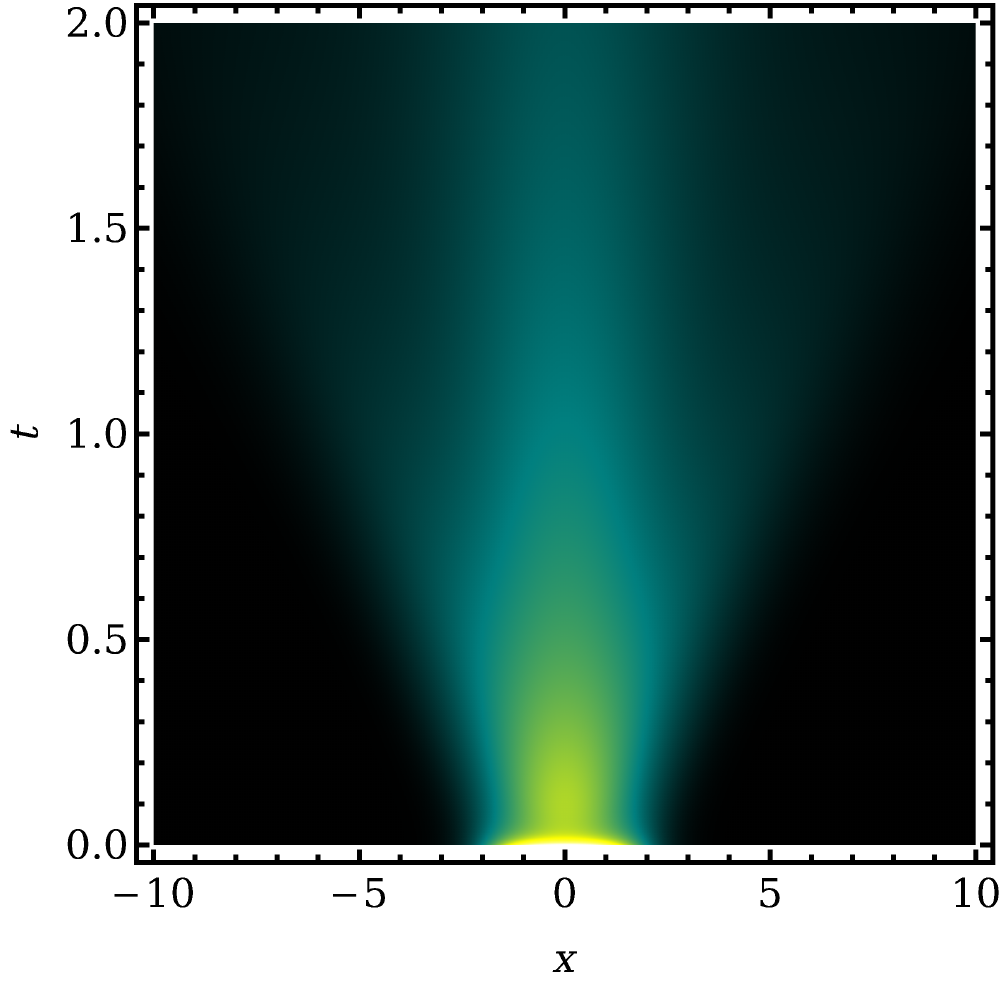}\\[1ex]
\raisebox{45ex}[0ex][0ex]{(b)}\includegraphics[width=0.75\columnwidth]{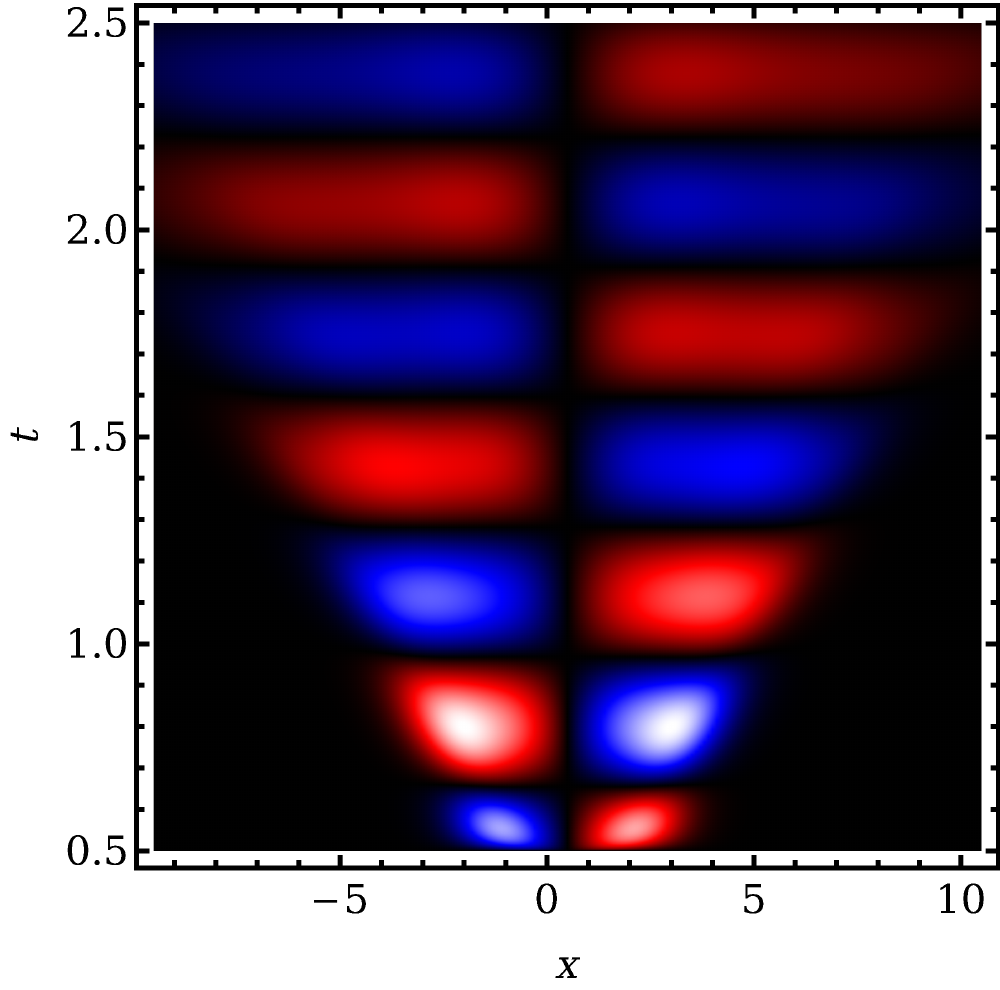}
\caption{Spatial distribution function of (a) the occupation number of triplet excitons and (b) the \textit{x} component of their spin polarization as functions of time in an applied magnetic field in the \textit{z} direction. The initial distribution of triplet excitons has helical spin polarization with the spin parallel (antiparallel) to $x$ for $k>0$ ($k<0$). The dimensionsless Larmor frequency is $\omega_L = 10$, the other parameters are the same as in Fig.~\ref{fig.NM.4}.
\label{fig.M.1}}
\end{figure}

In the previous examples, the spin of the triplet excitons does not matter. The initial singlet distribution is of course nonmagnetic and there is no mechanism that generates any spin polarization. To exhibit a characteristic magnetic effect, we consider an initial state with a distribution of \emph{triplet} excitons with momenta close to $\pm k_0$ and helical spin polarization: the spin is polarized parallel (antiparallel) to $x$ for $k>0$ ($k<0$). Except for the spin polarization, this distribution is similar to the one obtained by singlet fission in the previous examples. The occupation number $\rho_t$ of triplet excitons, shown in Fig.\ \ref{fig.M.1}(a), is independent of the applied magnetic field. The \textit{x} component of the spin polarization given by
\begin{equation}
s_t^x(x,t) = \int \frac{dk}{2\pi}\, \frac{1}{2} \Tr \widehat{S}_1 \widehat f_t(k,x,t)
\end{equation}
and plotted in Fig.\ \ref{fig.M.1}(b) shows two effects: First, due to the helical spin polarization of the initial distribution, triplet excitons moving to the right (left) have positive (negative) spin. Second, their spin precesses about the magnetic field, i.e., the \textit{z} direction, with a frequency given by the Larmor frequency $\omega_L$ in Eq.~(\ref{eq.example.LHStriplet}).

\section{Summary and conclusions}
\label{sec.concl}

The understanding of transport of charge carriers and excitons in organic semiconductors is crucial for applications. Semiclassical Boltzmann theory is a valuable tool for this since it is formulated in terms of physically intuitive quantities and processes, thereby simplifying the interpretation. In this paper, we have addressed the spin-conserving charge transport and scattering in a semiclassical framework, including the creation, decay, recombination, fission, and fusion of excitons, which are crucial for organic solar cells. However, our results are of more general interest for any system showing scattering of and reactions between spin-carrying particles, for example in spintronics and spin chemistry.

To start with, we have set up kinetic equations for charge carriers as well as for singlet and triplet excitons in uniform electric and magnetic fields, taking care to preserve $\mathrm{SU}(2)$ spin-rotation invariance. The case of triplet excitons is the most interesting in that these spin-$1$ quasiparticles permit and require a \emph{quadrupolar} contribution to their distribution function. We have obtained coupled equations for the concentration, spin-density, and quadrupole-density components. The generalization to nonuniform systems is conceptually straightforward.

In the second step, we have constructed $\mathrm{SU}(2)$-in\-va\-ri\-ant transition terms or scattering integrals describing disorder scattering, singlet-exciton generation, binding of electrons and holes into singlet and into triplet excitons, and exciton fusion, as well as the reverse processes. The occupation of final states enhances a process for bosons and suppresses it for fermions. These quantum effects are naturally included and lead to higher-order terms in the scattering integrals, which are relevant for large occupation numbers. It proved useful to employ two different representations of distribution functions, namely a multipole expansion and a description as matrix-valued functions acting on the spin Hilbert space. For the simpler processes, the multipole representation allows one to directly write down the $\mathrm{SU}(2)$-in\-va\-ri\-ant terms based on the transition terms under the (symmetry-breaking) assumption of all moments being parallel or antiparallel to the $z$ axis. The results can be used to describe real materials when realistic models for the bare transition rates $W$ are available. Moreover, our work serves as a guide on how to construct scattering integrals for complex transitions involving several quasiparticles with spin.

Future directions following from this work are quite clear. First, it will be necessary to derive models for the bare transition rates $W$ for real materials and link the equations to observables, in order to obtain quantitative descriptions. For the application to photovoltaic devices, it is important to note that the generation of excitons takes place at internal interfaces in organic blends, while the diffusion of (triplet) excitons is a bulk effect. Our framework offers two approaches to this problem: On long length scales relative to the typical scale of the phases in the blend, coarse graining leads to an effective-medium description with an effective exciton-generation rate. On the other hand, on shorter length scales---but still long compared to the single-molecule scale---the spatial structure of the phases can be modeled. The semiclassical approach is in principle suitable for this but of course relies on a good structural model.

Second, additional processes are relevant in certain materials. For example, bound states of two electrons and one hole or vice versa, i.e., trions, can exist \cite{Lam58,Gla20}. The formation and decay of trions can easily be described in our framework. Third, we have here assumed that polaronic effects, which are strong in organic materials, can be treated simply be renormalizing model parameters for charge carriers and excitons. This is certainly simplistic, in particular due to the broad distribution of timescales of phonons in organic materials. The connection of transport and excitonic processes with polaronic effects in a semiclassical framework is thus an important goal for the future.

\begin{acknowledgments}
The authors thank A. Knoll, K. Leo, F. Ortmann, and S. Reineke for stimulating discussions. C.\,T. acknowledges financial support by the Deutsche For\-schungs\-ge\-mein\-schaft through Collaborative Research Center SFB 1143, project A4, and the Cluster of Excellence on Complexity and Topology in Quantum Matter ct.qmat (EXC~2147).
\end{acknowledgments}

\appendix

\section{Kinetic equation for triplet excitons}
\label{app.kinetic.triplet}

In this Appendix, we derive kinetic equations for the concentration, spin, and quadrupolar coefficients of the nonequilibrium triplet-exciton distribution function in Eq.\ (\ref{eq:gt Q-ansatz}). We first introduce an alternative way to express the quadrupole term that simplifies calculations. Using the Cartesian quadrupole tensor operator defined by Eq.\ (\ref{eq:def_gmat}), the distribution function can be written as
\begin{equation}\label{eq:f_t G basis}
	\triplet = \mto \Snull + \ct \cdot \Sp  + \vb{a} \cdot \gmat \vb{b} .
\end{equation}
The decomposition contains 10 real coefficients but has only nine degrees of freedom, so we choose $|\vb{a}| = 1$ as an additional condition. We call relations such as Eq.\ (\ref{eq:f_t G basis}) \textit{matrix expansions}. Analogously, the deviation is written as
\begin{equation}\label{eq:gt G basis}
	\tripletg = \mtog \Snull + \ctg \cdot \Sp  + \atg \cdot \gmat \btg ,
\end{equation}
with $|\atg| = 1$.

Next, we evaluate the Boltzmann equation (\ref{eq:boltzmann f_T= f0,g}) in the matrix expansion,  
\begin{align}
	  & \frac{\partial}{\partial t}\, \mtog \Snull + \frac{\partial}{\partial t}\, \ctig \mspin{S}_i
	  + \frac{\partial}{\partial t}\, \big(\atgi_m \btgi_n \big)\, \gmatij_{mn} \nonumber\\
	  & \quad{} + \frac{1}{\hbar}\, \vb{\nabla}_{\vb{k}} \epsilon_t \cdot
	  \left[\vb{\nabla}_{\vb{r}} \mtog \Snull +  \vb{\nabla}_{\vb{r}}\ctig \mspin{S}_i
	    + \vb{\nabla}_{\vb{r}} \big(\atgi_m \btgi_n \big)\, \gmatij_{mn}\right] \nonumber\\
	  & \quad{} + \frac{g_\mathrm{tr} \mu_B}{2 \hbar}\, \varepsilon_{ijk} \btgi_j \ctkg \mspin{S}_i
	  + \frac{g_\mathrm{tr} \mu_B}{2 \hbar} \left(\vb{B} \times \atg  \right)_m \btgi_n  \gmatij_{mn}  \nonumber\\
	  & \quad{} + \frac{g_\mathrm{tr} \mu_B}{2 \hbar}\, \big(\vb{B} \times \btg \big)_m\, \atgi_n  \gmatij_{mn}
	    \nonumber \\
	 &= \scatpart{\depend}{t} ,
\end{align}
where $\gmatij_{mn}$ denote the components in Eq.\ (\ref{eq:def_gmat}) and $\varepsilon_{ijk}$ is the Levi-Civita symbol. We employ the Einstein summation convention. Equations for the number, spin, and quadrupole densities can be obtained using the relations
\begin{align}
	\big[ \widehat{S}_a, \widehat{S}_b \big] &= i \varepsilon_{abm} \widehat{S}_m , \\
	2\,\one_3 &= \widehat{S}_1^2 + \widehat{S}_2^2 + \widehat{S}_3^2 , \\
	\big[\gmatij_{ab}, \widehat{S}_c \big] &= i \varepsilon_{acm} \gmatij_{mb}
	  + i \varepsilon_{bcm} \gmatij_{ma} , \\
	\gmatij_{ab} \widehat{S}_c &= \frac{i}{2}\, \big(\varepsilon_{bcm}\gmatij_{am}
	  + \varepsilon_{acm}\gmatij_{bm} \big) \nonumber \\
	&\quad{} + \frac{1}{2}\, \big( \delta_{ac}\widehat{S}_b +  \delta_{cb}\widehat{S}_a \big)
	  - \frac{1}{3}\, \delta_{ab} \widehat{S}_c ,
\end{align}
with $a,b,c,m \in \{1,2,3\}$. After the Boltzmann equation has been solved with regard to Eq.\ (\ref{eq:f_t G basis}), we can convert the quadrupole coefficients to the form of the multipole expansion in Eq.\ (\ref{eq:f_t Q basis}) by the transformation
\begin{align}
	\qtx &= a_2b_3 + a_3b_2 , \\
	\qty &= a_3b_1 + a_1b_3 , \\
	\qtz &= a_1b_2 + a_2b_1 , \\
	\qtxy &= \frac{1}{2}(a_1b_1 -a_2b_2) , \\
	\qtzr &= - \frac{\sqrt{3}}{6} (a_1b_1 +a_2b_2 - 2a_3b_3) .  
\end{align}
Analogously to the density matrix, we expand the scattering term as
\begin{equation}
\widehat{I}_t
  = I^{n_t}\, \Snull + \vb{I}^{\vb{c}_t} \cdot \Sp + \vec{I}^{\,\vec{q}} \cdot \vec{\Q} .
\end{equation}
Using the differential operator $\mathcal{D}_t$ defined in Eq.\ (\ref{eq:def Dt}), we obtain Eqs.\ (\ref{eq:boltzmann triplet 0})--(\ref{eq:boltzmann triplet 8}).


\begin{thebibliography}{105}
\bibitem{note.smallpolaron}The term ``small polaron'' is somewhat confusing since it is also used for the logically independent limit of strong electron-vibration coupling, for the concept of self-trapped polarons, and for the case of thermally activated transport.
\bibitem{note.Haberkorn}Haberkorn \cite{Hab76} is concerned with the positivity of $\widehat\rho$, not with the Hermiticity, as erroneously stated in Ref.\ \cite{SWK11}.
\bibitem{note.soc}In the presence of spin-orbit coupling, the terms containing the velocity have to be symmetrized according to
$$
\vb{v} \cdot \Nabla_{\vb{r}} \widehat f
  \to \frac{1}{2} \left( \widehat{\vb{v}} \cdot \Nabla_{\vb{r}} \widehat f
  + \Nabla_{\vb{r}} \widehat f \cdot \widehat{\vb{v}} \right)
$$
etc.\ to ensure the Hermiticity of $\widehat f$.
\bibitem{note.nonlinear}We do not assume the dependence to be linear, i.e., the coefficient $c_{e0}$ may depend on the field strength $B$.
\bibitem{note.anticomm}The anticommutator is superfluous under the trace; it is introduced here for later convenience.
\end{thebibliography}
\end{document}